%% file: main.tex
\documentclass[a4paper,11pt]{article}
\pdfoutput=1
\usepackage{jheppub}
\makeatletter
\gdef\@fpheader{}
\makeatother
\usepackage[T1]{fontenc}
\usepackage{lmodern,microtype}
\usepackage{mathtools,bm,mathrsfs}
\usepackage{booktabs,array}
\usepackage{tikz}
\usetikzlibrary{arrows.meta,calc,decorations.markings}
\usepackage[nameinlink,noabbrev]{cleveref}

\crefformat{section}{section~#2#1#3}
\Crefformat{section}{Section~#2#1#3}
\crefformat{subsection}{section~#2#1#3}
\Crefformat{subsection}{Section~#2#1#3}
\crefformat{subsubsection}{section~#2#1#3}
\Crefformat{subsubsection}{Section~#2#1#3}
\crefformat{appendix}{appendix~#2#1#3}
\Crefformat{appendix}{Appendix~#2#1#3}
\AddToHook{cmd/appendix/before}{%
  \crefalias{section}{appendix}\crefalias{subsection}{appendix}}
\hypersetup{
  pdftitle={Out-of-time-order correlators with conformal boundary conditions},
  pdfkeywords={conformal boundary conditions, BTZ, de Sitter, OTOC, retarded response}
}
\newcommand{\dd}{\mathrm d}
\newcommand{\ii}{\mathrm i}
\newcommand{\Tr}{\operatorname{Tr}}
\newcommand{\sgn}{\operatorname{sgn}}
\newcommand{\Res}{\operatorname{Res}}
\newcommand{\cO}{\mathcal O}
\newcommand{\cB}{\mathcal B}
\newcommand{\cI}{\mathcal I}
\newcommand{\cD}{\mathcal D}
\newcommand{\cH}{\mathcal H}
\newcommand{\D}{\Delta}
\newcommand{\bp}{p_b}
\newcommand{\bv}{v_b}
\newcommand{\kb}{k_b}
\newcommand{\rb}{r_b}
\newcommand{\rh}{r_h}
\newcommand{\tb}{\widetilde\beta}
\newcommand{\Om}{\widetilde\Omega}
\newcommand{\Li}{\operatorname{Li}}
\newcommand{\hP}{\hat\Phi}%
\newcommand{\hr}{\hat\rho}%
\allowdisplaybreaks
\preprint{UT-WI-38-2026}
\title{Out-of-time-order correlators with conformal boundary conditions}

 \author[]{Hare Krishna}
 \affiliation[]{Weinberg Institute, Department of Physics,\\
   The University of Texas at Austin, Austin, TX 78712, USA}
 \emailAdd{hkrishna.phy@gmail.com}

\abstract{
We study out-of-time-order correlators (OTOCs) in three-dimensional
gravity with a finite timelike boundary. Fixing the conformal class of the
induced metric and the trace of the extrinsic curvature leaves the Weyl
factor of the boundary metric dynamical. In the axisymmetric sector of
rotating BTZ and de Sitter spacetimes, we compute the leading
gravitational contribution to the OTOC of two probe fields that couple
only to the reduced two-dimensional metric. In boundary conformal time
$u$, the OTOC has two growing terms, $e^{2\pi u/\widetilde\beta}$ and
$e^{2\pi\nu u/\widetilde\beta}$, where $\widetilde\beta$ is the inverse
temperature in this frame and $\nu=r_h/r_b$ is the ratio of the horizon
radius to the boundary radius. The first term gives the usual thermal
growth and dominates in the BTZ exterior. The second arises from a
classical instability of the boundary's Weyl mode and dominates in the
inner-horizon BTZ and de~Sitter patches for noninteger $\nu>1$.
The same instability appears in the time-ordered correlator.
For $\nu>1$, the factorization
estimate used to derive the chaos bound fails in this range.
}

\begin{document}
\maketitle
\flushbottom

\section{Introduction and summary}
\label{sec:summary}

How fast does a quantum system with many degrees of freedom forget a
small perturbation? Classically, the answer is a Lyapunov exponent. Its
quantum counterpart can be read off from the squared commutator
$-\langle[W(t),V(0)]^2\rangle_\beta$ of two
simple operators~\cite{LarkinOvchinnikov},
or equivalently from the out-of-time-order correlator (OTOC)~\cite{Kitaev2014,SS,MSS},
\begin{equation}
 \frac{\langle W(t)VW(t)V\rangle_\beta}{\langle WW\rangle_\beta\langle VV\rangle_\beta}
 \;\sim\;1-\frac1N\,e^{\lambda_Lt}.
 \label{eq:introcommutator}
\end{equation} Here $N$ counts the
degrees of freedom and the growth persists until the scrambling time
$t_*\sim\lambda_L^{-1}\log N$, when the perturbation has spread over
the whole system~\cite{HaydenPreskill,SekinoSusskind,Lashkari}. The
OTOC thus measures the scrambling of quantum
information~\cite{Hosur}. Using AdS/CFT, one can compute it using holographic methods. Black holes are the fastest scramblers allowed by
quantum mechanics~\cite{SekinoSusskind}. In a unitary theory with a
large hierarchy between dissipation and scrambling times the exponent
is bounded, $\lambda_L\leq2\pi/\beta$~\cite{MSS}, and horizons in
Einstein gravity saturate the bound.

This connection was established in several steps. Shenker and Stanford ~\cite{SS} showed that a small perturbation of the
thermofield double state~\cite{MaldacenaEternal} becomes a shock wave on the horizon of the dual black hole. The strength of this shock grows as $e^{2\pi t/\beta}$, and it destroys the correlations between
the two sides. Multiple
perturbations, perturbations localized in space, and string
corrections were treated in~\cite{SSmultiple,RSSlocalized,SSstringy}.
 String corrections lower the exponent below its Einstein value. Kitaev
proposed that a maximal exponent should be regarded as a signature of
a gravitational dual~\cite{Kitaev2014,Kitaev2015}.
Sachdev--Ye--Kitaev (SYK)
model~\cite{SachdevYe,Kitaev2015,MaldacenaStanford,KitaevSuh}
provides a solvable quantum mechanical system in which both maximal
and submaximal chaos can be studied.

On the field theory side, the same exponent follows from conformal symmetry and a large central charge. In two-dimensional CFTs with a sparse light spectrum, the Virasoro identity block controls the OTOC,
and in higher dimensions the stress tensor block plays the same
role~\cite{RobertsStanford,JMV,FitzpatrickKaplan,TuriaciVerlinde}. The exchange of higher-spin currents violates the bound, so that chaos
constrains the space of holographic CFTs~\cite{Perlmutter}. The
degrees of freedom responsible for the growth can be isolated in an
effective theory of reparametrization
modes~\cite{HaehlRozali,CotlerJensen,BlakeLeeLiu,GSS}. Thus $\lambda_L$ is a
sharp diagnostic of the gravitational sector: Einstein gravity gives
$2\pi/\beta$, and departures from this value signal stringy physics.

In gravity the exponent is computed in one of two ways. The first is the eikonal method. The perturbation created by $W$ is boosted relative to the one created by $V$ by a factor $e^{2\pi t/\beta}$. Its gravitational field is a Dray--'t~Hooft shock
wave~\cite{DraytHooft,tHooft,KVV}, and the OTOC is the overlap of the
states before and after the collision, with a phase proportional to
$G_N\,e^{2\pi t/\beta}$~\cite{SS,SSstringy}. The second method
applies in low dimensions, where gravity has no local excitations and
the interaction is carried by a boundary mode. In nearly-AdS$_2$
gravity, this is the Schwarzian
reparametrization~\cite{AlmheiriPolchinski,Jensen,MSY,EMV,MTV,LMTV},
and in three dimensions the boundary gravitons play the same
role~\cite{CotlerJensen,HaehlRozali}. For the rotating BTZ black hole
they give two rates, set by the left- and right-moving
temperatures~\cite{ReynoldsRoss,Stikonas,Poojary,JKY,MezeiSarosi,Craps}.
For cosmological horizons the picture is less settled. A shock wave at
a de~Sitter horizon produces a time advance instead of a
delay~\cite{Aalsma,AGH}, and the OTOC has been found to
oscillate~\cite{AGH} or to grow at twice the rate allowed by the
de~Sitter temperature~\cite{KolchmeyerLiu} (see
\cite{SusskindAddendum,CSTY,HZ,MNX,CuiKolchmeyer} for further
discussion).

In all of the AdS computations above the operators are placed on an
asymptotic boundary, where gravity is not dynamical. More generally,
one would like to describe gravity in a finite region, with the dual
data on a timelike surface at finite
distance~\cite{York86,BrownYork,AHH}. The obvious choice is to fix the
induced metric on this surface. This Dirichlet problem is related to the correspondence between cutoff AdS and $T\bar{T}$-deformed CFTs~\cite{MMV,KLM,HKST} correspondence. It has analogues for de~Sitter
space~\cite{Coleman,BanihashemiJacobson,SVVV,AnninosHarris} and for JT
gravity at finite cutoff~\cite{JTFiniteCutoff,StanfordYang}. In
Euclidean signature, the Dirichlet condition is not elliptic,
so there is no good perturbative expansion about a generic
saddle~\cite{Anderson08,WittenBC}, and in Lorentzian signature the
corresponding initial boundary value problem is not well
posed~\cite{AnAnderson21,AndradeDirichlet}. Fixing the conformal class of the induced metric and the trace of its extrinsic curvature gives elliptic boundary conditions in Euclidean signature ~\cite{York72,Anderson08,WittenBC}. In Lorentzian signature, well-posedness depends on the appropriate setting ~\cite{AnAnderson21,AnAnderson25,LiuReallSantosWiseman25}.

The thermodynamics of gravity with conformal boundary conditions has
been worked out in some
detail~\cite{AGMflat,Observatories,AAGMAdS4,LSW,BSSflat,BSSthermal,
NearExtremal,Liouville}. The entropy is given by the horizon area, and
an analogue of the Ryu--Takayanagi formula is discussed
in~\cite{Caceres:2026yvs}. At high temperature, the free energy has the
form expected of a local conformal theory in one less
dimension~\cite{BSSflat,BSSthermal}. Since only the conformal class of
the induced metric is fixed, its Weyl factor becomes a dynamical field
on the boundary, and the would-be dual is an ordinary local field
theory coupled to this mode~\cite{BSSthermal}. This mode affects the
correlation functions of matter on the boundary. The question we shall
address is how it affects the OTOC. Does a horizon seen from a finite
conformal boundary still scramble at the rate $2\pi/\beta$, and does
the Weyl mode produce a growth of its own?

We shall answer these questions in three-dimensional gravity. Specifically, we study the
rotating BTZ and Kerr--de~Sitter spacetimes with a timelike boundary,
which we call the wall, at $r=\rb$. For rotating BTZ, we consider two
patches, the exterior $r_+<r<\rb$ and the region $\rb<r<r_-$ inside
the inner horizon. In each case $\rh$ denotes the horizon at which the
Euclidean geometry caps off. We shall refer to the patches in which
the horizon surrounds the wall, namely the inner-horizon patch of BTZ
and the de~Sitter patch, as cosmic patches. Restricting to
configurations that do not depend on the angle and reducing along the
angular circle gives a two-dimensional dilaton gravity on a disk.

Our matter sector consists of two decoupled two-dimensional CFTs on
the disk, with primaries $V$ and $W$ of dimensions $\D_V$ and $\D_W$.
As in the Schwarzian analysis of~\cite{MSY}, these theories couple to
the reduced metric but not to the dilaton or to the Kaluza--Klein
gauge field. This choice defines our two-dimensional matter model (see
\cref{sec:matter}). We define $V$ and $W$ in the boundary conformal
frame, with metric
$-\dd u^2+\dd\varphi^2$, whose inverse temperature we denote by $\tb$.
The dimensionless Lorentzian and Euclidean times are
\begin{equation}
 T=\frac{2\pi u}{\tb},
 \qquad
 \tau=\frac{2\pi u_E}{\tb}\sim\tau+2\pi .
 \label{eq:summaryclock}
\end{equation}

Let $F(T)$ be the OTOC with the four operators equally spaced around
the thermal circle, normalized by the two gravitationally corrected
two-point functions (see \cref{sec:otoc-def}). At late times we find
\begin{equation}
 \boxed{
 F(T)=1+g \D_V\D_W\left[
 -\frac{8 r_h}{p_0}\,e^{T}
 +\frac{2 r_h(\nu-\bv)^2}{\bp\,\nu\sin(\pi\nu/2)}\,e^{\nu T}+  O(1)
 \right]+O(g^2),
 \qquad
 \nu=\frac{\rh}{\rb},
 }
 \label{eq:mainOTOC}
\end{equation}
where $g=G_3/\rh$ and $O(1)$ stands for terms of first order in $g$
that remain bounded as $T\to\infty$. This is the main result of this
paper. The coefficients are determined by the background. If
$e^{2\rho_0(x)}(\dd x^2+\dd\theta^2)$ is the reduced Euclidean metric,
with the wall at $x=0$ and the horizon at $x\to-\infty$, then
\begin{equation}
 p(x)=\frac{\dd r}{\dd x},\qquad
 \bp=p(0),\qquad
 p_0=\lim_{x\to-\infty}e^{-2x}p(x),\qquad
 \bv=\frac{\bp}{\rb}-\rho_0'(0),
 \label{eq:summarydefs}
\end{equation}
so that $r=\rh+\tfrac12p_0e^{2x}+\cdots$ near the horizon. Both $\bp$
and $p_0$ are positive in the exterior and negative in a cosmic patch.
Explicit expressions for all the backgrounds are given in
\cref{sec:explicit}.

The two terms in \eqref{eq:mainOTOC} have different physical origins.
The $e^{T}$ term is the familiar thermal growth of the Schwarzian
theory.  The $e^{\nu T}$ term is driven by the boundary
Weyl mode $\sigma=\delta\log\Phi_b$. For a wall at fixed $K$, this mode
obeys $\ddot\sigma=\nu^2\sigma$ and is therefore linearly unstable.
 Its mixing with the boundary reparametrization transmits the
instability to the matter correlator. The relation $\nu=\rh/\rb$
holds for every background considered here, with or without rotation.
At fixed horizon and wall radii, rotation changes only the
coefficients of the two terms.

\begin{table}[t]
 \centering
 \begin{tabular}{@{}lcccc@{}}
  \toprule
  Patch & Range & $e^{T}$ term & $e^{\nu T}$ term
  & Dominant rate in $u$\\
  \midrule
  BTZ exterior & $0<\nu<1$ & $-$ & $+$ & $2\pi/\tb$\\
  BTZ inner-horizon & $1<\nu<2$ & $+$ & $-$ & $2\pi\nu/\tb$\\
  dS cosmological & $1<\nu<2$ & $+$ & $-$ & $2\pi\nu/\tb$\\
  \bottomrule
 \end{tabular}
 \caption{Signs of the growing terms in
\eqref{eq:mainOTOC}. In the equally
spaced OTOC $F(T)$, the insertions $V,W,V,W$ lie at Euclidean angles
$0,\pi/2,\pi,3\pi/2$, respectively. Shifting both $W$ insertions by
$\delta$ with $|\delta|<\pi/2$ preserves the alternating order.
In $\operatorname{Re}F$, this shift multiplies the thermal and Weyl
coefficients by $\cos\delta$ and $\cos(\nu\delta)$, respectively.
The thermal term keep its sign, while the real part of the Weyl
term can reverse sign even for $1<\nu<2$. The growing term in the
time-ordered correlator $C(T)$ is positive in all three
rows. Table~\ref{tab:cosmic-otoc-signs} gives the cosmic-patch signs
for all noninteger $\nu>1$.}
 \label{tab:signs}
\end{table}

The signs of the two terms are summarized in \cref{tab:signs}. In the
BTZ exterior, the thermal term dominates and has the usual negative
coefficient. The leading exterior OTOC correction grows at the thermal rate \(2\pi/\widetilde\beta\) in boundary time \(u\) with a coefficient that depends on the position of the
wall. In the cosmic patches, the coefficients of both terms reverse
sign. For every noninteger \(\nu>1\), the Weyl term gives the leading late-time growth of the connected OTOC. Its sign depends on \(\nu\), as shown in table \ref{tab:cosmic-otoc-signs}. Thus the inner-horizon patch of BTZ
behaves like the de~Sitter patch, although it is locally AdS$_3$. In
both cases $\rh>\rb$, and the Weyl contribution grows faster than the
thermal one.

\textbf{Dual field theory perspective:-} For rotating BTZ, the two growth regimes can be understood within
the boundary theory. The proposed dual is a holographic CFT
coupled to timelike Liouville theory and deformed by a marginal
dressed $T\bar T$ operator~\cite{Liouville}. In the convention of Allameh et.al.\cite{Liouville}, the dressed \(T\bar T\) coupling is \(\lambda=16\pi G_3\ell\) and the Liouville cosmological constant is \(\mu=(K\ell-2)/\lambda\).  Writing
$c_K=c_{\rm eff}/c_m$ for the ratio of the effective central charge
governing the high-temperature entropy to the matter central charge gives $\nu^2=(1-c_K^2)/(1-\Om^2)$. At fixed couplings, the
angular potential $\Om$ determines which rate dominates. For
$|\Om|<c_K$ the thermal term leads. For $c_K<|\Om|<1$ the wall lies
inside the inner horizon, and Weyl mode gives the leading growth of the connected OTOC
for noninteger $\nu$. The boundary thermal generator remains timelike
in this range, and the semiclassical torus partition function of the
proposed dual reproduces the inner-horizon entropy
$2\pi r_-/(4G_3)$~\cite{Liouville}. A direct field-theory derivation
of the OTOC growth is left for future works.

For $\nu>1$ the leading connected contribution to the OTOC grows
faster than $e^{2\pi u/\tb}$. This does not by itself contradict the
chaos bound of~\cite{MSS}. The argument of~\cite{MSS} uses the
normalized, thermally regularized time-ordered four-point function
\begin{equation}
 C(T)=
 \frac{\Tr[\rho^{1/2}W(T)V\rho^{1/2}VW(T)]}{N_VN_W},
 \qquad
 N_O=\Tr(\rho^{1/2}O\rho^{1/2}O),
 \label{eq:introC}
\end{equation}
whose disconnected part is $1$. We denote the connected part by
$\delta C(T)=C(T)-1$. For a positive thermal density matrix and
Hermitian operators, the Cauchy--Schwarz inequality gives
$|F(T\pm i\pi/2)|\le C(T)$. The argument then requires
$C(T)\le1+\epsilon$ throughout the relevant time interval after
dissipation. The bound follows when the factorization error
$\epsilon$ is small compared with $1-F(T)$. In our calculation, the connected gravitational exchange gives
\begin{equation}
 C(T)=1+ g \D_V\D_W\,
 \frac{2 r_h(\nu-\bv)^2}{\bp\,\nu}
 \cot\frac{\pi\nu}{2}\,e^{\nu T}
 +O(g)+O(g^2),
 \label{eq:mainTOC}
\end{equation}
where the $O(g)$ remainder is bounded at late times. Thus for $\nu>1$
the connected time-ordered correlator grows at the same exponential
rate as the leading
OTOC. Hence, the factorization error cannot be neglected.

The signs in table \ref{tab:cosmic-otoc-signs} follow from
\eqref{eq:mainOTOC} and \eqref{eq:mainTOC}. In both cosmic patches
$p_0<0$, so the thermal term in $F-1$ is positive. Since $p_b<0$,
the Weyl term in $F-1$ has sign $-\sgn[\sin(\pi\nu/2)]$ and
changes sign across each even integer. In $C-1$ the same Weyl
coefficient acquires a factor of $\cos(\pi\nu/2)$, giving the sign
$-\sgn[\cot(\pi\nu/2)]$, which changes across every integer.


The rate $\nu$ comes from a classical instability. A wall held at
fixed $K$ runs away as $\frac{d^2}{d T^2}\sigma=\nu^2\sigma$, independently of the
matter (\cref{sec:weyl}). The matter correlators inherit this growth
in every ordering we have computed, including $C(T)$ and the retarded
commutator. The thermal term doesn't contribute to either of those two. Hence, we do
not interpret the $e^{\nu T}$ term as faster scrambling.  Our results hold at tree level, in a Gaussian expansion about
a Euclidean saddle whose quadratic action is not positive definite
(\cref{sec:gaussian}). So a positive thermal density matrix is not
established. We also exclude integer $\nu\geq2$, where the Weyl mode has a zero mode.
As the wall approaches the horizon of a fixed nonextremal background (\(\nu\to1\)), the two growing terms combine at fixed \(T\) to give \(-(4T+8)\,g\Delta_V\Delta_W\,e^T\). 
\begin{table}[t]
  \centering
  \small
  \setlength{\tabcolsep}{5pt}
  \renewcommand{\arraystretch}{1.2}
  \begin{tabular}{@{}lcccc@{}}
    \hline
    Range of \(\nu\) & \(e^T\) in \(F-1\) & \(e^{\nu T}\) in \(F-1\)
      & \(e^{\nu T}\) in \(C-1\) & \shortstack{Dominant rate\\in \(u\)} \\
    \hline
    \((4k+1,\,4k+2)\) & \(+\) & \(-\) & \(+\) & \(2\pi\nu/\tilde{\beta}\) \\
    \((4k+2,\,4k+3)\) & \(+\) & \(+\) & \(-\) & \(2\pi\nu/\tilde{\beta}\) \\
    \((4k+3,\,4k+4)\) & \(+\) & \(+\) & \(+\) & \(2\pi\nu/\tilde{\beta}\) \\
    \((4k+4,\,4k+5)\) & \(+\) & \(-\) & \(-\) & \(2\pi\nu/\tilde{\beta}\) \\
    \hline
  \end{tabular}
  \caption{Signs of the growing terms in the cosmic patch BTZ
    and Kerr-de Sitter patches for \(k=0,1,2,\ldots\) for fixed noninteger \(\nu=r_h/r_b>1\).
    The Weyl term \(e^{\nu T}\) dominates the OTOC \(F\) and the time-ordered  correlator \(C\). The entries for \(\nu>2\) follow by analytic continuation
    of the Euclidean mode sum. In both correlators the $V$ insertions
lie at Euclidean angles $0$ and $\pi$. The $W$ insertions lie at
$\pi/2$ and $3\pi/2$ in the equally spaced OTOC $F$, and at
$\epsilon$ and $\pi-\epsilon$ in the time-ordered correlator $C$.  }
  \label{tab:cosmic-otoc-signs}
\end{table}

The rest of the paper is organized as follows. \Cref{sec:geometry}
describes the backgrounds, the reduced action, and the matter sector.
In \cref{sec:action}, we derive the quadratic action for the two
boundary modes. The correlators, together with the squared commutator, are defined
and computed in \cref{sec:contours}, and evaluated on the individual
backgrounds in \cref{sec:explicit}. \Cref{sec:retarded} discusses
retarded responses. We conclude in
\cref{sec:discussion} with a discussion of our analysis.
\Cref{app:hessian,app:nonrot,app:GMS,app:TOC,app:fullyR} contain technical
derivations.

\section{Setup}
\label{sec:geometry}

In this section, we shall describe the backgrounds and the boundary
data, and reduce the theory on the angular circle to a two-dimensional
dilaton gravity probed by free scalars. We discuss the reduced gravitational action as well as the probe matter fields whose correlator we want to compute. 

\subsection{Backgrounds and boundary data}
\label{sec:backgrounds}

We shall consider three families of backgrounds: the exterior of a
rotating BTZ black hole, the region of the same spacetime inside its
inner horizon, and the cosmological patch of Kerr--de~Sitter space in
three dimensions. In each case we cut the spacetime off at a wall, on
which we impose conformal boundary conditions, and keep the region
between the wall and a horizon.

The rotating BTZ metric is
\begin{equation}
 \begin{aligned}
 \dd s_3^2
 &=-f(r)\dd t^2+\frac{\dd r^2}{f(r)}
   +r^2\left(\dd\phi-\frac{j}{r^2}\dd t\right)^2,\\
 f(r)
 &=\frac{(r^2-r_+^2)(r^2-r_-^2)}{\ell^2r^2},
 \qquad
 j=\frac{r_+r_-}{\ell},
 \end{aligned}
 \label{eq:BTZ}
\end{equation}
where $\phi\sim\phi+2\pi$, $r_+>r_-\geq0$, and the angular momentum is
$J=j/(4G_3)$. We place a conformal \textit{wall} at $r=\rb$ and keep
one of the two regions (see fig \ref{fig:patches} for the placement of wall)
\begin{equation}
 \begin{array}{lll}
 \text{exterior:}
 & r_+\leq r\leq\rb,
 & \rh=r_+,\quad \varsigma=+1,\\[2pt]
 \text{cosmic:}
 & \rb\leq r\leq r_-,
 & \rh=r_-,\quad \varsigma=-1.
 \end{array}
 \label{eq:patches}
\end{equation}
Here $\rh$ denotes the horizon that bounds the region, and $\varsigma$
is the sign of the radial component of the outward unit normal at the
wall, $n=\varsigma\sqrt{f}\,\partial_r$. In both cases the normal
points away from the horizon, towards larger $r$ in the exterior and
towards smaller $r$ in the cosmic patch. The cosmic patch exists only
for $r_->0$. Both regions are parts of the rotating BTZ spacetime,
and are therefore locally AdS$_3$~\cite{Liouville}.

The cosmological patch of Kerr--dS$_3$~\cite{BMS} has a metric of the
same form \eqref{eq:BTZ}, with
\begin{equation}
 f(r)=\frac{(\rh^2-r^2)(r^2+a^2)}{\ell^2r^2},
 \qquad
 j=\frac{\rh a}{\ell},
 \label{eq:dS}
\end{equation}
where $\ell$ now denotes the de~Sitter radius and $a$ is the rotation
parameter. We keep the region $\rb\leq r\leq\rh$ between the wall and
the cosmological horizon, so that $\varsigma=-1$. For $a=0$ and
$\rh=\ell$ this is the static patch.

In all three cases the surface gravity of the horizon at $r=\rh$ is
\begin{equation}
 \kappa=\frac{|f'(\rh)|}{2}
 =\begin{cases}
   (r_+^2-r_-^2)/(\ell^2\rh), & \text{BTZ},\\[2pt]
   (\rh^2+a^2)/(\ell^2\rh), & \text{de Sitter}.
  \end{cases}
 \label{eq:kappa}
\end{equation}

To continue to Euclidean signature, we use the angle $\phi-jt/\rh^2$,
which co-rotates with the horizon, and set $t=-\ii t_E$ with $j$ kept
real. The three-dimensional metric is then complex, but the
two-dimensional metric obtained by reducing on the angular circle is
real. This metric is smooth at $r=\rh$ provided $t_E$ has period
$2\pi/\kappa$. In disk coordinates it takes the form
\begin{equation}
 \dd s_2^2
 =f\,\dd t_E^2+\frac{\dd r^2}{f}
 =e^{2\rho_0(x)}(\dd x^2+\dd\theta^2),
 \qquad
 \theta=\kappa t_E,\quad
 x=\varsigma\kappa\int_{\rb}^{r}\frac{\dd R}{f(R)},\quad
 e^{2\rho_0}=\frac{f}{\kappa^2}.
 \label{eq:disk}
\end{equation}
Thus the reduced Euclidean geometry is a disk, with the wall at $x=0$
and the horizon at its centre, $x\to-\infty$.

Two functions of the background will appear repeatedly. They are
\begin{equation}
 p(x)=\frac{\dd r}{\dd x}=\frac{\varsigma f}{\kappa},
 \qquad
 k(x)=\frac{\dd\rho_0}{\dd x}=\frac{\varsigma f'(r)}{2\kappa},
 \qquad
 p'=2kp,
 \label{eq:pk}
\end{equation}
where the last relation follows from $e^{2\rho_0}=f/\kappa^2$. A
subscript $b$ will denote the value of a quantity at the wall, so that
$\bp=p(0)$ and $\kb=k(0)$. Primes on functions of $x$ denote
derivatives with respect to $x$, while $f'(r)$ denotes a derivative
with respect to $r$.

On the wall it is convenient to use the coordinates
\begin{equation}
 u=\frac{\sqrt{f_b}}{\rb}\,t,
 \qquad
 \varphi=\phi-\frac{j}{\rb^2}\,t,
 \label{eq:wall-coordinates}
\end{equation}
in which the induced metric takes the form
$\rb^2(-\dd u^2+\dd\varphi^2)$. In these coordinates, smoothness at the
horizon becomes the pair of identifications
\begin{equation}
 (u_E,\varphi)\sim(u_E+\tb,\varphi-\ii\tb\Om),
 \qquad
 (u_E,\varphi)\sim(u_E,\varphi+2\pi),
 \label{eq:boundary-thermal-identification}
\end{equation}
with
\begin{equation}
 \tb=\frac{2\pi\sqrt{f_b}}{\kappa\rb}\equiv2\pi b,
 \qquad
 \Om=\frac{j\rb}{\sqrt{f_b}}
       \left(\frac{1}{\rh^2}-\frac{1}{\rb^2}\right).
 \label{eq:clock}
\end{equation}
Thus the conformal class of the boundary torus is specified by
$(\tb,\Om)$. The dimensionless time of \eqref{eq:summaryclock} is
$T=u/b=\kappa t$, and on the background the disk angle $\theta$
coincides with $\tau=u_E/b$.

The remaining boundary data is the trace of the extrinsic curvature
of the wall,
\begin{equation}
 K=\varsigma\left(
       \frac{f_b'}{2\sqrt{f_b}}+\frac{\sqrt{f_b}}{\rb}
     \right),
 \qquad\text{equivalently}\qquad
 Kb\rb^2=\bp+\rb\kb.
 \label{eq:K}
\end{equation}
Together with $(\tb,\Om)$ it labels the ensemble.

\subsection{Reduction on the angular circle}
\label{sec:reduction}

\begin{figure}[tbp]
 \centering
 \input{figures/patches.tex}
 \caption{ \textbf{Left}: Penrose diagram of the rotating BTZ in
 $(t,r)$ plane. The outer horizons
 $r=r_+$ cross at the bifurcation point $B_+$, and the inner horizons
 $r=r_-$ at $B_-$. Region~I ($r>r_+$) ends on the asymptotic boundary,
 region~III ($0<r<r_-$) ends at $r=0$, and II and II$'$ are the black-
 and white-hole regions $r_-<r<r_+$. The diagram repeats vertically.
 Arrows point towards increasing $r$. The exterior patch (blue) lies in
 I between $r_+$ and the wall, and the cosmic patch (red) lies in III
 between the wall and $r_-$. Wall positions are shown by dark line.
\textbf{ Right}: the same backgrounds in the reduced theory, where $\Phi=r$ and
 $(\nabla\Phi)^2=f(\Phi)$. The horizons are the zeros of $f$, and the
 slope there is $f'(\rh)=U(\rh)$. The two patches are the intervals
 between a horizon and the wall on which $f>0$, with $U>0$ on the
 exterior patch and $U<0$ on the cosmic patch. For $r_-<\Phi<r_+$,
 $f<0$ and no static wall exists. The plot uses $(r_+,r_-)=(2\ell,\ell)$,
 with walls at $\rb=3\ell$ ($\nu=2/3$) and $\rb=3\ell/4$ ($\nu=4/3$).}
 \label{fig:patches}
\end{figure}
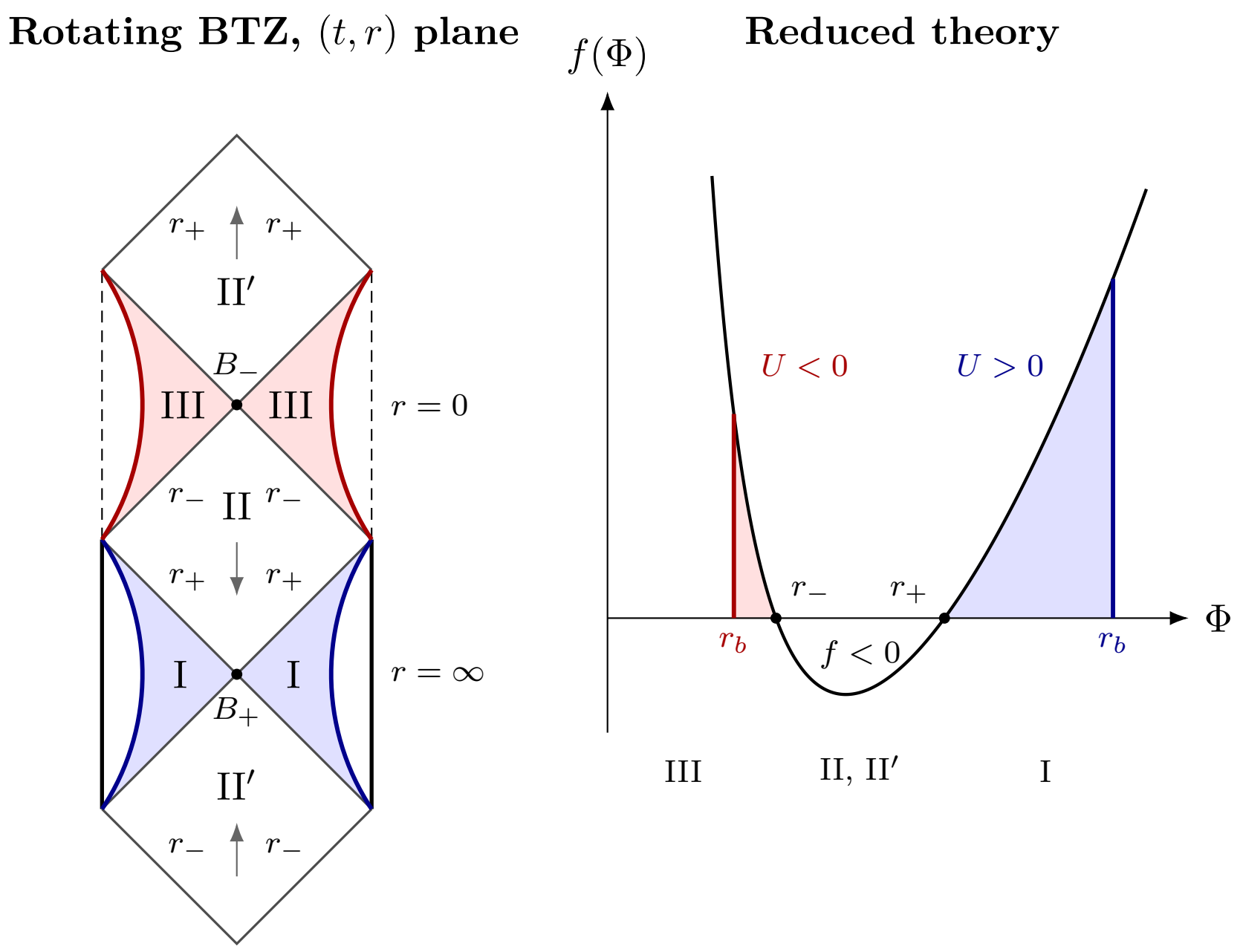

We now restrict to configurations that do not depend on $\phi$, and
write the three-dimensional metric in Kaluza--Klein form (see fig \ref{fig:patches} for BTZ reduction),
\begin{equation}
 \dd s_3^2
 =g_{\mu\nu}\dd x^\mu\dd x^\nu
  +\Phi^2\left(\dd\phi+A_\mu\dd x^\mu\right)^2.
 \label{eq:KK}
\end{equation}
On the backgrounds of \cref{sec:backgrounds}, $\Phi=r$ and
$A=-(j/r^2)\,\dd t$. The Einstein--Hilbert action reduces to a
two-dimensional dilaton gravity coupled to the gauge field $A_\mu$,
whose kinetic term is proportional to $\Phi^3F^2$. Since no matter
field is charged under $A_\mu$, the Maxwell equation
$\nabla_\mu(\Phi^3F^{\mu\nu})=0$ implies that $\Phi^3{*F}$ is
constant. On the backgrounds $F_{rt}=2j/r^3$, and this constant is
proportional to the angular momentum.

We can therefore integrate out the gauge field at fixed $J$. This
requires a boundary Legendre transform that trades the fixed holonomy
for a fixed flux, and the net effect of the gauge field is an extra
term in the dilaton potential. With Dirichlet conditions on the
two-dimensional metric and dilaton, the action is
\begin{equation}
 I_D
 =-\frac{1}{8G_3}\int_M\dd^2x\,\sqrt{g}\,
       \bigl(\Phi R+U(\Phi)\bigr)
   -\frac{1}{4G_3}\int_{\partial M}\dd\ell\,\Phi k_2,
 \qquad
 U(\Phi)=-2\Lambda\Phi-\frac{2j^2}{\Phi^3},
 \label{eq:ID}
\end{equation}
where $k_2$ is the extrinsic curvature of the boundary curve, and
$\Lambda=-1/\ell^2$ for BTZ and $\Lambda=+1/\ell^2$ for de~Sitter. On
the stationary solutions $U(r)=f'(r)$.

Conformal boundary conditions fix instead the ratio
$\dd\ell/\Phi_b=\dd u_E$, the holonomy, and the three-dimensional
trace
\begin{equation}
 K=k_2+\frac{n^\mu\nabla_\mu\Phi}{\Phi}.
 \label{eq:K-reduced}
\end{equation}
The corresponding action differs from \eqref{eq:ID} by a boundary
term,
\begin{equation}
 I_K=I_D+\frac{K}{8G_3}\int_{\partial M}\dd\ell\,\Phi,
 \label{eq:action}
\end{equation}
in which $K$ is now part of the boundary data. On the stationary
solutions \eqref{eq:K-reduced} reproduces \eqref{eq:K}.

The ensemble at fixed $\Om$ is obtained by summing over $J$,
\begin{equation}
 Z(\tb,\Om,K)=\sum_J e^{\tb\Om J_\partial}\,Z_J(\tb,K),
 \qquad
 J_\partial=\varsigma J,
 \label{eq:ensemble}
\end{equation}
where $Z_J$ is computed with the action $I_K$. The sign $\varsigma$
appears because the charge conjugate to $\Om$ is measured with the
outward normal at the wall. In the saddle point approximation the sum
extremizes $I_K-\tb\Om J_\partial$ with respect to $J$, and this
reproduces the relation \eqref{eq:clock} between $\Om$ and the
background.

Finally, in conformal gauge,
$\dd s_2^2=e^{2\rho(x,\theta)}(\dd x^2+\dd\theta^2)$ with
$-\infty<x\leq0$, the action \eqref{eq:ID} takes the form
\begin{equation}
 I_D
 =-\frac{1}{4G_3}\int_{-\infty}^{0}\dd x
                    \int_0^{2\pi}\dd\theta\,
 \left(
   \partial\Phi\cdot\partial\rho
   +\frac12 e^{2\rho}U(\Phi)
 \right)
 +I_{\rm cap},
 \qquad
 I_{\rm cap}=-\frac{2\pi\Phi_h}{4G_3},
 \label{eq:confaction}
\end{equation}
where the dot product uses the flat metric and $\Phi_h$ is the value
of the dilaton at the centre of the disk. The term $I_{\rm cap}$ comes
from the smooth cap at the centre when the curvature term is
integrated by parts. On the saddle it equals minus the horizon
entropy. The equations of motion of $\rho$ and $\Phi$ are
\begin{equation}
 (\partial_x^2+\partial_\theta^2)\Phi=e^{2\rho}U(\Phi),
 \qquad
 (\partial_x^2+\partial_\theta^2)\rho=\frac12e^{2\rho}U'(\Phi).
 \label{eq:mixed}
\end{equation}
They must be supplemented by the constraints \footnote{ The first equation in \eqref{eq:mixed} is the trace of the metric equation,
since varying $\rho$ gives $\delta_\rho g_{ab}=2g_{ab}\,\delta\rho$.
The two traceless constraints follow by varying the action \eqref{eq:ID}
with respect to $g^{zz}$ and $g^{\bar z\bar z}$ before fixing conformal gauge.
In this gauge $g_{zz}=0$ and $\Gamma^z_{zz}=2\partial_z\rho$, so the
$zz$ constraint is
\begin{equation}
 \nabla_z\nabla_z\Phi
 =\partial_z^2\Phi-2\partial_z\rho\,\partial_z\Phi=0,
 \qquad z=x+\ii\theta,\nonumber
\end{equation}
and the $\bar z\bar z$ constraint is its complex conjugate.}, which are the traceless
components of the metric equation,
\begin{equation}
 \partial_z^2\Phi-2\partial_z\rho\,\partial_z\Phi=0,
 \qquad
 z=x+\ii\theta,
 \label{eq:bulk-constraint}
\end{equation}
together with the complex conjugate equation.

We shall \emph{not} impose the constraints in the bulk. Instead we
shall solve \eqref{eq:mixed} for given boundary values of $\Phi$ and
$\rho$, and evaluate $I_K$ on the solution. The result is an action
for the boundary fields, and we shall see in \cref{sec:hessian} that
the constraints arise as one of its equations of motion. This is the
same logic that leads to the Schwarzian action in JT gravity, where
the boundary curve is kept off shell until its own equation of motion
is imposed.
\subsection{Matter probes}
\label{sec:matter}

To probe the gravitational interaction we add two free massless
scalars on the disk,
\begin{equation}
 I_{\rm matter}
 =\frac12\int_M\dd^2x\,\sqrt g\,
 \left[(\partial\chi_V)^2+(\partial\chi_W)^2\right].
 \label{eq:probe-matter}
\end{equation}
They do not couple to $\Phi$ or to $A_\mu$. This is the matter
coupling of~\cite{MSY}. \footnote{Scalar matter that does not couple to the dilaton also appears
in the work of Almheiri and Polchinski~\cite{AlmheiriPolchinski}.
Ghosh, Maxfield and Turiaci~\cite{Ghosh:2019rcj} discuss how BTZ scalar
matter reduces to this form in the near-horizon region of a near-extremal
rotating BTZ black hole.}
We impose Dirichlet conditions $\chi_V|_{\partial M}=v(u_E)$ and
$\chi_W|_{\partial M}=w(u_E)$, and define the boundary operators $V$
and $W$ by differentiating the generating functional with respect to
the sources $v$ and $w$. Thus $V$ and $W$ are the normal derivatives
of the scalars at the wall, and they have dimensions $\D_V=\D_W=1$.
The sources are set to zero after differentiation, so the
gravitational saddles are not modified. We shall keep $\D_V$ and
$\D_W$ general, since the formulas below apply to the boundary
primaries of any two decoupled CFTs that couple to gravity in the
same way.

Since the two sectors are decoupled, on a fixed geometry the
four-point function factorizes,
\begin{equation}
 \langle V_1V_2W_3W_4\rangle_g
 =\langle V_1V_2\rangle_g\,\langle W_3W_4\rangle_g.
 \label{eq:factorization}
\end{equation}
The two factors depend on the same metric, and integrating over its
fluctuations correlates them. At leading order in $G_3$ the connected
four-point function is obtained by expanding each two-point function
to first order in the boundary modes and contracting the two
variations with the gravitational propagator. We keep the matter
central charges and the operator dimensions fixed as $G_3/\rh\to0$.
The propagator is then the one computed from the pure gravity action,
and matter loops correct it only at higher order.

Let us now discuss how \eqref{eq:probe-matter} is related to matter in
three dimensions. A minimally coupled massless scalar $\Psi$ that does
not depend on $\phi$ reduces as
\begin{equation}
 \frac12\int\dd^3x\,\sqrt{g_3}\,(\partial\Psi)^2
 =\frac12\int_M\dd^2x\,\sqrt g\,
       \frac{\Phi}{\Phi_\star}\,(\partial\chi)^2,
 \qquad
 \chi=\sqrt{2\pi\Phi_\star}\,\Psi,
 \label{eq:reduced-scalar-matter}
\end{equation}
where $\Phi_\star$ is an arbitrary reference radius that fixes the
normalization of $\chi$. With $\Phi_\star=\rh$, the background factor
$\Phi_0/\Phi_\star=r/\rh$ is close to unity throughout the
patch when $|\rb-\rh|\ll\rh$. In this regime, \eqref{eq:probe-matter}
approximates the free propagation of the reduced scalar on the fixed
geometry. Expanding \eqref{eq:reduced-scalar-matter} to first order in
$\delta\Phi$ also produces a vertex proportional to
$(\delta\Phi/\rh)(\nabla\chi)^2$, which can contribute to the
connected four-point function at the same order in $G_3$ as the
exchange computed below. We use \eqref{eq:probe-matter} to define our
two-dimensional probes at any admissible wall radius. Extending the
calculation to the $s$-wave of a minimally coupled three-dimensional
scalar requires evaluating this additional vertex, which we leave to
future work.


\section{The quadratic boundary action}
\label{sec:action}

Let $u_E$ be Euclidean time in the boundary conformal frame, with
thermal period $\tb$. We use the dimensionless boundary time
$\tau=2\pi u_E/\tb$, so that $\tau\sim\tau+2\pi$.

In this section we compute the quadratic action for the two boundary
modes. The Weyl mode $\sigma(\tau)$ rescales the proper circumference
of the spatial circle at the wall by $e^{\sigma(\tau)}$. A point on
the wall labelled by $\tau$ has angular coordinate
$\theta(\tau)=\tau+\xi(\tau)$ on the Euclidean disk. The shift
$\xi(\tau)$ is the reparametrization mode.

Since the constraints are not imposed in the bulk,
the linearized solution contains at each frequency a constant $c_n$
that measures their violation. The action depends on $\xi$ only
through $c_n$, so the equation of motion of $\xi$ restores the
constraint. And integrating out $\xi$ leaves a local action for
$\sigma$ that grows at the rate $\nu=\rh/\rb$ on every background we
consider. The four-point function needs only the propagator $C_n$ of
$\xi$, given in \eqref{eq:CnGeneral}. Its continuation in $n$ has a
pole at $n=1$, which produces the thermal growth of the OTOC. A
second pole at $n=\nu$ comes from the Weyl mode.

\subsection{Boundary degrees of freedom}
\label{sec:boundary-response}

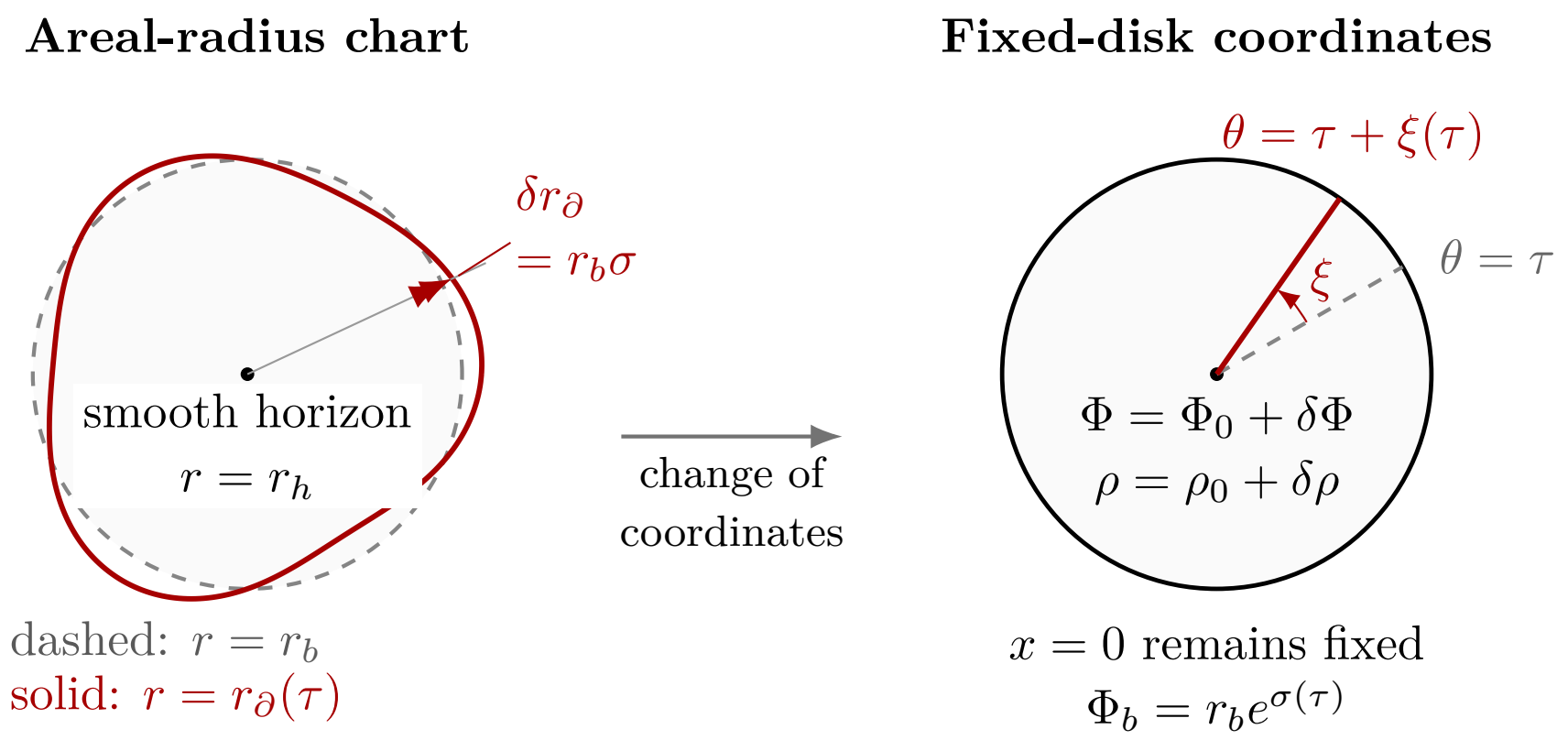
\begin{figure}[tbp]
 \centering
 \begin{tikzpicture}[
   >=Latex, font=\small,
   base/.style={draw=black!48,dashed,line width=.8pt},
   pert/.style={draw=red!65!black,line width=1.1pt}]

  \begin{scope}[shift={(-3.5,0)}]
   \node[font=\small\bfseries] at (0,2.45)
     {Areal-radius chart};

   \fill[black!2] (0,0) circle (1.55);
   \draw[base] (0,0) circle (1.55);

   \draw[pert,smooth,samples=181,domain=0:360,variable=\a]
     plot ({(1.55+.14*cos(3*\a)+.06*sin(2*\a))*cos(\a)},
           {(1.55+.14*cos(3*\a)+.06*sin(2*\a))*sin(\a)});

   \fill (0,0) circle (1.4pt);
   \node[align=center,fill=white,inner sep=2pt]
     at (0,-.52) {smooth horizon\\$r=\rh$};

   \draw[black!40] (0,0)--(25:1.9);
   \draw[<->,red!65!black,line width=.8pt]
     (25:1.55)--(25:{1.55+.14*cos(75)+.06*sin(50)});
   \draw[red!65!black] (25:1.68)--(1.9,.95);
   \node[anchor=west,red!65!black,align=left]
     at (1.8,1.05) {$\delta r_{\partial}$\\$=\rb\sigma$};

   \node[anchor=north west,black!65] at (-1.85,-1.65)
     {dashed: $r=\rb$};
   \node[anchor=north west,red!65!black] at (-1.85,-2.02)
     {solid: $r=r_{\partial}(\tau)$};
  \end{scope}

  \draw[->,black!55,line width=.8pt] (-.8,-.45)--(.8,-.45);
  \node[align=center,font=\footnotesize] at (0,-.92)
    {change of\\coordinates};

  \begin{scope}[shift={(3.5,0)}]
   \node[font=\small\bfseries] at (0,2.45)
     {Fixed-disk coordinates};

   \fill[black!2] (0,0) circle (1.55);
   \draw[line width=.9pt] (0,0) circle (1.55);
   \fill (0,0) circle (1.4pt);

   \draw[base] (0,0)--(30:1.55);
   \draw[pert] (0,0)--(55:1.55);
   \draw[->,red!65!black] (30:.75)
     arc[start angle=30,end angle=55,radius=.75];

   \node[red!65!black] at (42:1.02) {$\xi$};
   \node[anchor=west,black!60] at (30:1.7) {$\theta=\tau$};
   \node[anchor=south,red!65!black] at (55:1.72)
     {$\theta=\tau+\xi(\tau)$};

   \node[align=center] at (0,-.55)
     {$\Phi=\Phi_0+\delta\Phi$\\$\rho=\rho_0+\delta\rho$};
   \node[anchor=north,align=center] at (0,-1.68)
     {$x=0$ remains fixed\\$\Phi_b=\rb e^{\sigma(\tau)}$};
  \end{scope}
 \end{tikzpicture}

 \caption{The same off-shell wall configuration in two coordinate systems
on the reduced Euclidean disk of an exterior patch. The spatial
Kaluza-Klein circle is suppressed. In the areal-radius chart (left),
the dashed curve is the reference wall at $r=r_b$, and the solid curve
is the displaced wall at $r=r_b e^{\sigma(\tau)}$. The indicated
displacement is $\delta r_{\partial}=r_b\sigma$ to first order.
In conformal coordinates (right), the wall stays at $x=0$, its radius
is encoded in the boundary dilaton $\Phi_b(\tau)=r_b e^{\sigma(\tau)}$,
and $\theta(\tau)=\tau+\xi(\tau)$ gives the disk angle corresponding
to boundary-clock time $\tau$.}
 \label{fig:boundary-fluctuations}
\end{figure}

We describe the fluctuations in conformal gauge, and choose the
coordinates so that the wall stays at $x=0$:
\begin{equation}
 \dd s_2^2=e^{2\rho(x,\theta)}(\dd x^2+\dd\theta^2),
 \qquad
 \Phi=\Phi_0(x)+\delta\Phi(x,\theta),
 \qquad
 \rho=\rho_0(x)+\delta\rho(x,\theta),
 \label{eq:bulk-fluctuations}
\end{equation}
with $\Phi_0(x)=r(x)$ and $\rho_0$ as in \eqref{eq:disk}. In these
coordinates, the wall does not move. What fluctuates are the boundary
values of the fields and the relation between the disk angle $\theta$
and the boundary clock. In a chart that uses the areal radius $\Phi$
as the radial coordinate, the same configurations describe a wall at a
fluctuating position $R(\tau)=\rb e^{\sigma(\tau)}$. The two
descriptions are shown in \cref{fig:boundary-fluctuations}.

Conformal boundary conditions require the induced metric on the
wall to be
\begin{equation}
 \dd s_{\partial}^2
 =\rb^2e^{2\sigma(\tau)}\left(\dd u_E^2+\dd\varphi^2\right),
 \label{eq:boundary-conformal-metric}
\end{equation}
with the identifications \eqref{eq:boundary-thermal-identification}
and an arbitrary function $\sigma(\tau)$. Comparing with the
Kaluza--Klein form \eqref{eq:KK} gives
\begin{equation}
 \Phi_b(\tau)=\rb e^{\sigma(\tau)},
 \qquad
 \dd\ell=\Phi_b\,\dd u_E=b\Phi_b\,\dd\tau,
 \label{eq:cbc-line-element}
\end{equation}
so the Weyl mode rescales the angular circle and the line element of
the reduced boundary by the same factor. The second boundary field is
the disk angle as a function of the clock,
\begin{equation}
 \theta(\tau)=\tau+\xi(\tau).
 \label{eq:boundary-map}
\end{equation}
This is the analogue of the reparametrization mode of the Schwarzian
theory. In a boundary correlator, we hold the insertion times $\tau_i$ fixed.
For each configuration of $\xi$, the insertion at $\tau_i$ lies on the
wall $x=0$ at disk angle\footnote{For each allowed configuration of \(\xi\), the map \(\tau\mapsto\theta(\tau)=\tau+\xi(\tau)\) is one-to-one modulo \(2\pi\). The wall matching condition gives \(\theta'=1+\xi'>0\), and periodicity of \(\xi\) gives \(\theta(\tau+2\pi)=\theta(\tau)+2\pi\). Thus \(\theta\) goes around the circle exactly once. Varying \(\xi\) changes which disk angle corresponds to a fixed insertion time \(\tau_i\).} $\theta(\tau_i)=\tau_i+\xi(\tau_i)$.

The two fields determine the boundary values of $\Phi$ and $\rho$. Along the wall $x=0$, the metric gives $\dd\ell=e^{\rho(0,\theta(\tau))}\dd\theta$ with $\dd\theta=(1+\xi'(\tau))\dd\tau$, and \eqref{eq:cbc-line-element} gives the same line element as $\dd\ell=b\rb e^{\sigma(\tau)}\dd\tau$.
\begin{equation}
 e^{\rho(0,\theta(\tau))}\left(1+\xi'(\tau)\right)
 =b\rb e^{\sigma(\tau)}.
 \label{eq:exact-boundary-matching}
\end{equation}
Since $e^{\rho_0(0)}=b\rb$, to first order this gives
\begin{equation}
 \delta\Phi_b=\rb\sigma,
 \qquad
 \delta\rho_b=\sigma-\xi'.
 \label{eq:bdryfields}
\end{equation}
We shall also need the exact relation
\eqref{eq:exact-boundary-matching} at second order, in
\cref{sec:hessian}. Here and below primes on $\sigma$ and $\xi$ denote
derivatives with respect to $\tau$. Dirichlet boundary conditions
correspond to $\sigma=0$, with $\xi$ still free.

\subsection*{The radial problem}
\label{sec:radial}

We expand $\sigma=\sum_n\sigma_ne^{\ii n\tau}$ and
$\xi=\sum_n\xi_ne^{\ii n\tau}$. Different frequencies decouple at
quadratic order. We first consider $n\geq2$, and return to the modes
with $|n|\leq1$ at the end of this subsection. For $n\geq2$ we write
\begin{equation}
 \delta\Phi=\hP_n(x)e^{\ii n\theta},
 \qquad
 \delta\rho=\hr_n(x)e^{\ii n\theta},
 \qquad
 \hP_n(0)=\rb\sigma_n,
 \quad
 \hr_n(0)=\sigma_n-\ii n\xi_n,
 \label{eq:radial-ansatz}
\end{equation}
where the boundary values follow from \eqref{eq:bdryfields}. Negative
frequencies are obtained by complex conjugation. Linearizing
\eqref{eq:mixed} about the background gives
\begin{equation}
 \begin{aligned}
 (\partial_x^2-n^2)\hP_n
 &=e^{2\rho_0}\left(U'\hP_n+2U\hr_n\right),\\
 (\partial_x^2-n^2)\hr_n
 &=e^{2\rho_0}\left(\tfrac12U''\hP_n+U'\hr_n\right),
 \end{aligned}
 \label{eq:linearized-radial-eom}
\end{equation}
where $U$ and its derivatives are evaluated at $\Phi_0(x)$.\footnote{For any potential, this coupled system can be reduced to quadratures
once the background is known. The reason is that the combination
\begin{equation}
 E_n\equiv\hP_n'-k\hP_n-p\hr_n
 \label{eq:En}
\end{equation}
satisfies the decoupled equation $(\partial_x^2-n^2)E_n=0$, as a
consequence of the background equations (see \cref{app:hessian}).}

Regularity at the centre of the disk then requires
\begin{equation}
 \hP_n'-k\hP_n-p\hr_n=c_ne^{nx}.
 \label{eq:constraint}
\end{equation}
The integration constant $c_n$ measures the violation of the
linearized constraint, since
\begin{equation}
 \delta\!\left(\partial_z^2\Phi
 -2\partial_z\rho\,\partial_z\Phi\right)
 =n\,c_n\,e^{nz}.
 \label{eq:constraint-violation}
\end{equation}
The equations \eqref{eq:mixed} ensure that this expression is
holomorphic on the disk, but they do not require it to vanish for
generic boundary data.

Using \eqref{eq:constraint} to eliminate $\hr_n$, we can integrate the
remaining equation once more. Evaluating the result at the wall gives
(see \cref{app:hessian})
\begin{equation}
 \hr_n(0)=\frac{n+\kb}{\bp}\,\hP_n(0)+c_nD_n,
 \qquad
 D_n=\frac1{\bp}-4n\cI_n,
 \qquad
 \cI_n=\int_{-\infty}^{0}\frac{e^{2nx}}{p(x)}\,\dd x.
 \label{eq:DtN}
\end{equation}
The integral $\cI_n$ contains all the nonlocal dependence on the
interior. Near the horizon $p(x)\simeq p_0e^{2x}$ for some constant
$p_0$, so $\cI_n$ converges for $ n>1$, and its analytic
continuation has a simple pole at $n=1$.  Substituting the boundary values
\eqref{eq:radial-ansatz} into \eqref{eq:DtN} we get
\begin{equation}
 c_nD_n=a_n\sigma_n-\ii n\xi_n,
 \qquad
 a_n=1-\frac{\rb(n+\kb)}{\bp}=-\frac{\rb}{\bp}(n-\bv),
 \label{eq:response-boundary-data}
\end{equation}
where $\bv=\bp/\rb-\kb$, as defined in \eqref{eq:summarydefs}.

The modes with $|n|\leq1$ need separate treatment. The
reparametrizations $\xi_0$ and $\xi_{\pm1}$ generate the conformal
automorphisms of the disk and are pure gauge. The corresponding Weyl
modes $\sigma_0$ and $\sigma_{\pm1}$ are physical, but they do not
contribute to the exchange. The matter vertex couples only to $\xi$,
and since different frequencies decouple at quadratic order, these
modes do not mix with any mode that couples to matter.

\subsection*{The Hessian}
\label{sec:hessian}

The quadratic action for $\sigma$ and $\xi$ receives three
contributions. The first is the second variation of $I_D$ on the
linearized solution, which reduces to the boundary term
\begin{equation}
 -\frac1{8G_3}\int_0^{2\pi}\dd\theta\,
 \left(
   \delta\Phi_b\,\partial_x\delta\rho_b
   +\delta\rho_b\,\partial_x\delta\Phi_b
 \right).
 \label{eq:dirichlet-quadratic-response}
\end{equation}
The second comes from the relation between the boundary coordinate
$\tau$ and the bulk coordinate $\theta$. When the boundary data are
expressed in terms of $\theta$ using
\eqref{eq:exact-boundary-matching}, the boundary values acquire
second-order corrections. These contribute through the first
variation of $I_D$. The third is the boundary term in
\eqref{eq:action}, which takes a simple form in the coordinate $\tau$,
\begin{equation}
 I_K-I_D
 =\frac{Kb\rb^2}{8G_3}\int_0^{2\pi}e^{2\sigma(\tau)}\,\dd\tau.
 \label{eq:boundary-legendre-term}
\end{equation}
The terms linear in $\sigma$ cancel by \eqref{eq:K}, as they should,
since the background solves the equations of motion at this value of
$K$. Combining the three quadratic contributions we find (see
\cref{app:hessian})
\begin{equation}
 I^{(2)}=\frac{2\pi}{8G_3}\sum_{|n|\geq2}
 \left[H_0(n)\,\sigma_{-n}\sigma_n
 -\frac{(a_{|n|}\sigma_{-n}+\ii n\xi_{-n})
        (a_{|n|}\sigma_n-\ii n\xi_n)}{D_{|n|}}
 \right],
 \label{eq:quadratic-completed-square}
\end{equation}
where
\begin{equation}
 H_0(n)=-\frac{\rb^2}{\bp}\,(n^2-\nu^2),
 \qquad
 \nu^2=\frac{\bp^2}{\rb^2}-\frac{\bp\kb}{\rb}-k_b'+\kb^2.
 \label{eq:nuGeneral}
\end{equation}
It is convenient to collect the two fields into
$X_n=(\sigma_n,\xi_n)^{\mathsf T}$ and write
\begin{equation}
 I^{(2)}=\frac{2\pi}{8G_3}
      \sum_{|n|\geq2}X_{-n}^{\mathsf T}H_nX_n,
 \qquad
 H_n=
 \begin{pmatrix}
  H_0(n)-a_n^2/D_n & \ii n a_n/D_n\\
  -\ii n a_n/D_n & -n^2/D_n
 \end{pmatrix}.
 \label{eq:Hessian}
\end{equation}
Here the matrix is given for $n\geq2$, and $H_{-n}=H_n^{\mathsf T}$.
Its determinant is $\det H_n=-n^2H_0(n)/D_n$.
The reparametrization equation enforces the linearized bulk constraint
discussed in \cref{sec:reduction}.\footnote{For $n\geq2$, varying with
respect to $\xi_{-n}$ gives $(H_nX_n)_2=-\ii n c_n=0$ by
\eqref{eq:response-boundary-data}. With $c_n=0$, the $\sigma_{-n}$
equation reduces to $H_0(n)\sigma_n=0$.}

We have checked \eqref{eq:Hessian} in two ways. For $j=0$ the dilaton
can be integrated out exactly, leaving an action for a curve in the
hyperbolic plane or on the sphere, and its quadratic expansion
reproduces \eqref{eq:Hessian} (see \cref{app:nonrot}). Setting
$\sigma=0$ leaves only the reparametrization kernel $-n^2/D_n$. We
shall see in \cref{sec:propagators} that for nonrotating BTZ this
reduces to the Schwarzian kernel when the wall is far from the
horizon.

\subsection*{The Weyl mode and its growth rate}
\label{sec:weyl}

Integrating out $\xi$ gives a local action for the Weyl mode \footnote{Integrating out \(\xi_n\) in Eq. \eqref{eq:quadratic-completed-square} gives the following quadratic action for \(\sigma_{\geq2}(\tau)=\sum_{|n|\geq2}\sigma_n e^{in\tau}\). The physical Weyl modes with \(n=0,\pm1\) require separate treatment and do not contribute to the matter exchange computed below.},
\begin{equation}
 I^{(2)}_{\sigma}
 =-\frac{\rb^2}{8G_3\bp}\int_0^{2\pi}
       \left[(\sigma')^2-\nu^2\sigma^2\right]\dd\tau.
 \label{eq:WeylAction}
\end{equation}
In the exterior $\bp>0$, the kinetic term has the wrong sign. This
is the conformal factor problem. In
the cosmic and de~Sitter patches $\bp<0$. Continued to real time, $\tau\to\ii T$, \eqref{eq:WeylAction}
gives $\dd^2\sigma/\dd T^2=\nu^2\sigma$. Thus $\nu$ is the growth rate
of the Weyl mode in units of $2\pi/\tb$.

To evaluate $\nu$ we substitute \eqref{eq:pk} into
\eqref{eq:nuGeneral}. This gives
\begin{equation}
 \kappa^2\nu^2
 =\left[
    \frac{f'^2}{4}-\frac{ff''}{2}
       -\frac{ff'}{2r}+\frac{f^2}{r^2}
   \right]_{r=\rb}
 =\begin{cases}
  (r_+^2-r_-^2)^2/(\ell^4\rb^2),&\text{BTZ},\\[2pt]
  (\rh^2+a^2)^2/(\ell^4\rb^2),&\text{de Sitter}.
 \end{cases}
 \label{eq:identity}
\end{equation}
In both cases the right-hand side equals $\kappa^2\rh^2/\rb^2$, and
therefore
\begin{equation}
 \nu=\frac{\rh}{\rb}.
 \label{eq:nu}
\end{equation}
The same growth rate follows from the Lorentzian motion of the wall.
Since three-dimensional gravity has no local gravitational degrees of
freedom, the bulk geometry remains stationary at fixed mass and angular
momentum. For a wall of areal radius $R(\mathsf t)$, the trace of its
extrinsic curvature is
\begin{equation}
 K=\varsigma\left[
  \frac{\ddot R+\tfrac12 f'(R)}{\sqrt{f(R)+\dot R^2}}
  +\frac{\sqrt{f(R)+\dot R^2}}{R}
 \right],
 \label{eq:moving-wall-K}
\end{equation}
where dots denote derivatives with respect to the proper time
$\mathsf t$ along the wall in the reduced geometry.

Let $K_{\rm stat}(R)$ denote the value of $K$ for a static wall of
radius $R$ in the same bulk geometry. Linearizing the boundary
condition $K=K_{\rm stat}(\rb)$ about $R=\rb$, with
$\delta R=\rb\sigma$, gives
\[
 0=\frac{\delta K}{\rb}
   =K_{\rm stat}'(\rb)\,\sigma
    +\frac{\varsigma}{\sqrt{f_b}}\,\ddot\sigma .
\]
Using the background relation
$\dd\mathsf t=(\sqrt{f_b}/\kappa)\,\dd T$, we obtain
\begin{equation}
 \frac{\dd^2\sigma}{\dd T^2}=\nu^2\sigma,
 \qquad
 \nu^2=-\frac{\varsigma f_b^{3/2}}{\kappa^2}\,K_{\rm stat}'(\rb),
 \label{eq:lorentzian-weyl-frequency}
\end{equation}
in agreement with \eqref{eq:identity}.

\subsection{Propagators}
\label{sec:propagators}

With the normalization of the Gaussian integral fixed by
\eqref{eq:Hessian}, the propagators are
$\langle X_nX_{-n}^{\mathsf T}\rangle=(2G_3/\pi)H_n^{-1}$, where
\begin{equation}
 H_n^{-1}=
 \begin{pmatrix}
  H_0(n)^{-1} & \ii a_n/[nH_0(n)]\\
  -\ii a_n/[nH_0(n)] &
     -D_n/n^2+a_n^2/[n^2H_0(n)]
 \end{pmatrix}.
 \label{eq:inverse-hessian}
\end{equation}
In particular, for $n\geq2$,
\begin{align}
 \langle\sigma_n\sigma_{-n}\rangle
  &=-\frac{2G_3\bp}{\pi\rb^2(n^2-\nu^2)},
  \label{eq:sigprop}\\
 C_n\equiv\langle\xi_n\xi_{-n}\rangle
  &=\frac{2G_3}{\pi n^2}
     \left[
       4n\cI_n-\frac1{\bp}
       -\frac{(n-\bv)^2}{\bp(n^2-\nu^2)}
     \right],
  \label{eq:CnGeneral}
\end{align}
and $C_{-n}=C_n$. The four-point function requires only the
reparametrization propagator $C_n$. Its analytic continuation in $n$
has two poles in the right half-plane, one at $n=1$ inherited from
$\cI_n$, and one at $n=\nu$ coming from the Weyl mode. At large $n$
the expansion
\begin{equation}
 \cI_n=\frac{1}{2n\bp}+\frac{\kb}{2n^2\bp}+O(n^{-3})
 \label{eq:In-large-n}
\end{equation}
shows that the leading terms in $C_n$ cancel, leaving
\begin{equation}
 C_n=\frac{4G_3}{\pi\rb\,n^3}+O(n^{-4}).
 \label{eq:UV}
\end{equation}
This falls off more slowly than the $n^{-4}$ behaviour of the
Schwarzian propagator \eqref{eq:dirichlet-Schwarzian-limit} below,
which holds only for $\nu|n|\ll1$.

\paragraph{Dirichlet boundary conditions:-}
For Dirichlet boundary conditions, the Weyl factor is fixed, $\sigma=0$,
and only $\xi$ fluctuates. Its quadratic kernel is the $\xi\xi$ entry
of \eqref{eq:Hessian}, so that
\begin{equation}
 H_n^{(D)}=-\frac{n^2}{D_n},
 \qquad
 C_n^{(D)}=\frac{2G_3}{\pi n^2}\left(4n\cI_n-\frac1{\bp}\right).
 \label{eq:dirichlet-hessian}
\end{equation}
Allowing the Weyl factor to fluctuate changes the propagator by
\begin{equation}
 C_n-C_n^{(D)}
 =-\frac{2G_3}{\pi\bp}\,\frac{(n-\bv)^2}{n^2(n^2-\nu^2)}.
 \label{eq:CBC-Dirichlet-comparison}
\end{equation}
This correction carries the Weyl pole at $n=\nu$, and leaves the pole
at $n=1$ unchanged.

For nonrotating BTZ, the potential is linear, $U=2\Phi/\ell^2$, and
\eqref{eq:ID} reduces to JT gravity with a boundary at finite
distance~\cite{JTFiniteCutoff}. In this case everything can be computed
in closed form. Defining $q=(1-\nu)/(1+\nu)$, we have
$p=4\rh qe^{2x}/(1-qe^{2x})^2$, and the integral $\cI_n$ can be done
explicitly (see \eqref{eq:nonrot-radial-integral}). This gives
\begin{equation}
 D_n=-\frac{(1+n\nu)^2}{\rh(1-\nu^2)(n^2-1)},
 \qquad
 H_n^{(D)}=\frac{\rh(1-\nu^2)\,n^2(n^2-1)}{(1+\nu|n|)^2}.
 \label{eq:finite-JT-dirichlet-kernel}
\end{equation}
For $\nu|n|\ll1$ the kernel reduces to $\rh n^2(n^2-1)$. This is the
quadratic kernel obtained by expanding the Schwarzian action
\begin{equation}
 I_{\mathrm{Sch}}
 =-\frac{\rh}{4G_3}\int_0^{2\pi}
   \left\{\tan\frac{\theta(\tau)}2,\tau\right\}\dd\tau
 \label{eq:Schwarzian-action-comparison}
\end{equation}
about $\theta=\tau$~\cite{MSY}. In the same limit $\bv=-\nu\to0$, so
the correction \eqref{eq:CBC-Dirichlet-comparison} vanishes. Both
propagators therefore approach
\begin{equation}
 C_n,\ C_n^{(D)}\longrightarrow
 \frac{2G_3}{\pi\rh}\,\frac1{n^2(n^2-1)}.
 \label{eq:dirichlet-Schwarzian-limit}
\end{equation}

With rotation, the potential is no longer linear, and the reduced
theory is not JT gravity. JT gravity is recovered in the
near-extremal throat, where $U\simeq8(\Phi-r_e)/\ell^2$ with
$r_e^2=j\ell$. The expressions \eqref{eq:CnGeneral} and
\eqref{eq:dirichlet-hessian} remain valid for the full potential.

\paragraph{Relation to the Schwarzian-Liouville action:- }
Galante et.al.~\cite{NearExtremal} derived a boundary
action for the circle reduction of AdS$_3$ gravity with conformal
boundary conditions, valid near the asymptotic boundary. Their action
is the sum of a Schwarzian term for the boundary reparametrization and
a Liouville-type term for the logarithm of the boundary dilaton.
Our fields $\xi$ and $\sigma$ describe fluctuations of these two
variables. In \cref{app:GMS},  we show that for nonrotating BTZ the
limit $\nu\to0$ at fixed $n$ gives
\begin{equation}
 H_n\longrightarrow\rh
 \begin{pmatrix}
 -n^2 & 0\\
 0 & n^2(n^2-1)
 \end{pmatrix},
 \label{eq:GMS-Hessian-limit}
\end{equation}
in agreement with the quadratic expansion of their action. The mixing between $\sigma$ and
$\xi$, and the shift from $n^2$ to $n^2-\nu^2$ in the Weyl propagator,
are effects of the finite radius of the wall. They are absent from the
leading asymptotic action.


\section{The four-point function}
\label{sec:contours}

In this section we compute the four-point function at first
order in $g=G_3/\rh$. The operators couple only to the
reparametrization mode through the linearized Schwarzian bilocal. So
the connected four-point function is a single exchange of $\xi$ with
propagator $C_n$. Its late-time growth is controlled by the two poles
of $C_n$ at $n=1$ and $n=\nu$. The time-ordered correlator $C(T)$
grows with the second of these rates, while the squared commutator
vanishes at this order. Throughout this section
$\tau=u_E/b$ and $T=u/b$ are related by $\tau=\ii T$.



\subsection*{Integration contour and zero modes}
\label{sec:gaussian}
For the quadratic action \eqref{eq:quadratic-completed-square}, we
integrate the mode amplitude along the real axis in each positive
eigendirection of the Hessian and along the imaginary axis in each
negative eigendirection. This choice makes the Euclidean Gaussian
convergent in every nonzero eigendirection and gives the propagators
in \cref{sec:propagators}. \\

This prescription defines the correlators as a formal expansion about
the saddle. It does not determine the global contour of the
gravitational path integral. It also doesn't establish that the
correlators are expectation values in a positive thermal state. We
leave both questions open.\\
We omit the pure-gauge reparametrizations $\xi_0$ and $\xi_{\pm1}$
from the mode sum. When $\nu$ equals an integer $N\geq2$, the Weyl
kernel $H_0(N)$ vanishes, giving a physical zero mode in the
$|n|=N$ sector. The Gaussian integral over its amplitude then
diverges, so we work away from these values of $\nu$.

\subsection*{Operators and the bilocal vertex}
\label{sec:bilocal}

We label operator insertions by the boundary clock time $\tau$. In the
conformal gauge of \cref{sec:boundary-response} the reduced metric is
$g=e^{2\rho(x,\theta)}(\dd x^2+\dd\theta^2)$, and the insertion at
$\tau$ lies on the wall $x=0$ at angle $\theta(\tau)=\tau+\xi(\tau)$.
Since $\sqrt g\,g^{\mu\nu}=\delta^{\mu\nu}$ in these coordinates, the
free-scalar action \eqref{eq:probe-matter} becomes
\begin{equation}
 I_{\rm matter}
 =\frac12\sum_{A=V,W}\int_{-\infty}^{0}\dd x\int_0^{2\pi}\dd\theta\,
 \left[(\partial_x\chi_A)^2+(\partial_\theta\chi_A)^2\right].
 \label{eq:flat-matter}
\end{equation}
The scalar fields thus obey the flat Laplace equation on the
half-cylinder, with Dirichlet data at the wall and regularity at the
disk centre $x\to-\infty$.

Let $\cO_\theta$ denote a boundary primary in the conformal frame
with boundary metric $\dd\theta^2$. For each free scalar, the operator
conjugate to its Dirichlet source has dimension $\D=1$.
We keep $\D$ general, as in \cref{sec:matter}, and normalize the
angular two-point function to $[4\sin^2(\theta_{12}/2)]^{-\D}$,
where $\theta_{12}=\theta_1-\theta_2$.

The angular metric pulls back to $\theta'(\tau)^2\dd\tau^2$.
By the transformation law for a primary of dimension $\D$, the
operator in the clock frame $\dd\tau^2$ is therefore
\begin{equation}
 \cO(\tau)=[\theta'(\tau)]^\D\,\cO_\theta(\theta(\tau)).
 \label{eq:clock-primary}
\end{equation}
Taking the matter expectation value at fixed geometry gives the
Schwarzian bilocal
\begin{equation}
 \cB_\D(\tau_1,\tau_2)
 \equiv\langle\cO(\tau_1)\cO(\tau_2)\rangle_g
 =\left[
 \frac{(1+\xi'_1)(1+\xi'_2)}
      {4\sin^2\!\bigl((\tau_{12}+\xi_1-\xi_2)/2\bigr)}
 \right]^\D ,
 \label{eq:bilocal}
\end{equation}
where $\tau_{12}=\tau_1-\tau_2$, $\xi_i=\xi(\tau_i)$, and
$\xi'_i=\xi'(\tau_i)$. Expanding to first order in $\xi$ at fixed
insertion times gives
\begin{equation}
 L_{12}\equiv\delta\log\cB_\D
 =\D\left[
 \xi'_1+\xi'_2-\cot\frac{\tau_{12}}2\,(\xi_1-\xi_2)
 \right].
 \label{eq:vertex}
\end{equation}
This vertex vanishes for the three gauge modes $\xi=1,e^{\pm\ii\tau}$.
Since \eqref{eq:bilocal} depends only on $\xi$, the Weyl mode $\sigma$
enters the leading connected four-point function only through its
mixing with $\xi$ in the Hessian \eqref{eq:Hessian}.

\subsection{The OTOC}
\label{sec:otoc-def}

We normalize the four-point function by the two two-point functions,
each computed with its gravitational corrections \footnote{At this order, the normalization cancels the self-contractions within
each bilocal. By \eqref{eq:UV}, the coincident contraction
$\langle[\xi'(\tau)]^2\rangle=\sum_{|n|\geq2}n^2 C_n$ diverges
logarithmically. Expanding $[\theta'(\tau)]^\D=(1+\xi'(\tau))^\D$
at each endpoint gives a total contribution
$\D(\D-1)\langle[\xi'(\tau)]^2\rangle$ to the fractional correction
of the two-point function. Its coefficient vanishes for the free-scalar
probes with $\D=1$. For general $\D$, this divergence is independent
of the endpoint separation and can be absorbed into a multiplicative
renormalization of $\cO$. With a common regulator, it cancels directly
between the numerator and denominator of the normalized four-point
function, leaving $\mathcal M(12;34)$ finite at order $G_3$ for
separated insertions.}. When
$\langle\cB_V\cB_W\rangle/
(\langle\cB_V\rangle\langle\cB_W\rangle)$
is expanded to first order in $G_3$, we have
\begin{equation}
 \mathcal M(12;34)=\langle L_{12}L_{34}\rangle.
 \label{eq:Mdef}
\end{equation}

For the OTOC, we place the two pairs at antipodal points of the
Euclidean circle and continue the position of the second pair (see fig. \ref{fig:OTOcontour}, a),
\begin{equation}
 V:(0,\pi),\qquad
 W:(z,z+\pi),\qquad
 z=\frac{\pi}{2}+\ii T,
 \label{eq:OTOpositions}
\end{equation}
so that at $T=0$ the operators alternate around the circle. In
operator language, this is
\begin{equation}
 F(T)=\frac{\Tr[yVyW(T)yVyW(T)]}{N_VN_W},
 \qquad
 y=\rho^{1/4},\quad
 \rho=Z^{-1}e^{-\tb(H-\Om J_\partial)},
 \label{eq:traceF}
\end{equation}
with $N_O=\Tr(\rho^{1/2}O\rho^{1/2}O)$ and $W(T)\equiv W(u=bT)$. Here
the trace is a shorthand for the ordering of the insertions along the
contour. As explained in \cref{sec:gaussian}, we do not assume that a
positive $\rho$ exists. The operators are neutral,
$[J_\partial,V]=[J_\partial,W]=0$, so the angular potential plays no
role in the ordering. Note that the continuation in
\eqref{eq:OTOpositions} must be done after the sum over Fourier modes. The individual terms grow as $\cosh(nT)$.

The second configuration (see \cref{fig:OTOcontour}, b) we need is
\begin{equation}
 V:(0,\pi),\qquad
 W:(\epsilon+\ii T,\pi-\epsilon+\ii T),
 \qquad \epsilon\to0^+ ,
 \label{eq:Cpositions}
\end{equation}
with the limit taken after the continuation. Here the two pairs are
not interleaved: the two $W$ insertions are adjacent on the circle,
between the two $V$ insertions. In operator language this is the
thermally regularized time-ordered correlator
\begin{equation}
 C(T)=\frac{\Tr[\rho^{1/2}W(T)V\rho^{1/2}VW(T)]}{N_VN_W}.
 \label{eq:traceC}
\end{equation}
When $\rho$ is a positive density matrix and $V$, $W$ are Hermitian,
$C(T)$ is the squared Hilbert--Schmidt norm of $A=yW(T)Vy$, while
$F(T+\ii\pi/2)$ is $\Tr(A^2)/(N_VN_W)$ (see \cref{sec:norm}). The chaos bound~\cite{MSS} assumes that $C(T)$ stays
close to its factorized value $1$ after the dissipation time. Then
$|F|\leq1+\epsilon$ on both edges, and analyticity in the strip
limits the growth rate of $1-F$ to $2\pi/\tb$ in the boundary time
$u$.

\input{figures/contours_oto.tex}

\subsection*{The exchange kernel}
\label{sec:fourpoint}

In Fourier space $L_{12}=\D\sum_nb_n(1,2)\xi_n$, with
\begin{equation}
 b_n(1,2)=
 \ii n(e^{\ii n\tau_1}+e^{\ii n\tau_2})
 -\cot\frac{\tau_{12}}2\,(e^{\ii n\tau_1}-e^{\ii n\tau_2}),
 \label{eq:bn}
\end{equation}
and therefore
\begin{equation}
 \mathcal M(12;34)
 =\D_V\D_W\sum_{|n|\geq2}C_n\,b_n(1,2)\,b_{-n}(3,4).
 \label{eq:master}
\end{equation}
With $C_n$ given by \eqref{eq:CnGeneral} this is the complete
tree-level exchange. Since $C_n\sim n^{-3}$, the sum converges for
separated Euclidean points.

To continue to real time, we first resum. All Wightman orderings can be
obtained from the generating function $\cH(z)=\sum_{n\geq2}C_nz^n$. Then the Euclidean propagator of $\xi$ is
\begin{equation}
 G_\xi(\tau)
 =\lim_{\eta\to0^+}
 \left[\cH(e^{-\eta+\ii\tau})+\cH(e^{-\eta-\ii\tau})\right].
 \label{eq:Green}
\end{equation}
And $\mathcal M$ follows by acting with the derivatives in
\eqref{eq:vertex}. The sum can be done in closed form. 
The result is
\begin{align}
 \cH(z)={}&
 \frac{8G_3}{\pi}
 \int_{-\infty}^{0}\frac{L(e^{2x}z)}{p(x)}\,\dd x
 -\frac{2G_3}{\pi\bp}
 \left[
 \frac{2\bv}{\nu^2}L(z)
 +\left(1-\frac{\bv^2}{\nu^2}\right)P_2(z)
 +\frac{(\nu-\bv)^2}{2\nu^3}\Psi_\nu(z)
 -\frac{(\nu+\bv)^2}{2\nu^3}\Psi_{-\nu}(z)
 \right],
 \label{eq:Hfunction}
\end{align}
where
\begin{equation}
 L(z)=-\log(1-z)-z,\qquad
 P_2(z)=\Li_2(z)-z,\qquad
 \Psi_\alpha(z)=z^2\Phi_{\rm L}(z,1,2-\alpha)
   =\sum_{n=2}^{\infty}\frac{z^n}{n-\alpha},
 \label{eq:Hfunction-defs}
\end{equation}
and $\Phi_{\rm L}$ is the Lerch transcendent. These functions have a
branch cut along $z>1$, and the ordering of the operators determines
on which side of the cut each term is evaluated.

\subsection*{The equally spaced OTOC}
\label{sec:equally-spaced}

For antipodal pairs the cotangents in the vertices vanish, so that
$b_n(0,\pi)=\ii n[1+(-1)^n]$ and
$b_{-n}(z,z+\pi)=-\ii ne^{-\ii nz}[1+(-1)^n]$, and only even modes
contribute. Writing $F=1+\delta F+O(g^2)$, we get
\begin{equation}
 \delta F(z)
 =8\D_V\D_W\sum_{m=1}^{\infty}(2m)^2C_{2m}\cos(2mz).
 \label{eq:antipodal-exchange}
\end{equation}
For the rational part of $C_n$, we need the sums
\begin{equation}
 S_\alpha(z)
 =\sum_{m=1}^{\infty}\frac{\cos(2mz)}{2m-\alpha}
 =\frac14\left[
 w\,\Phi_{\rm L}(w,1,1-\alpha/2)
 +w^{-1}\Phi_{\rm L}(w^{-1},1,1-\alpha/2)
 \right],
 \qquad w=e^{2\ii z},
 \label{eq:Sa}
\end{equation}
with $\alpha=\pm\nu$, and $\sum_{m\geq1}\cos(2mz)=-1/2$. For
the radial part
\begin{equation}
 \sum_{m=1}^{\infty}2m\,e^{4mx}\cos(2mz)
 =\frac{e^{4x+2\ii z}}{(1-e^{4x+2\ii z})^2}
 +\frac{e^{4x-2\ii z}}{(1-e^{4x-2\ii z})^2}.
 \label{eq:radial-sum}
\end{equation}
On the trajectory $z=\pi/2+\ii T$ the exponentials are negative,
$e^{4x\pm2\ii z}=-e^{4x\mp2T}\equiv y_\pm$, and the integral over $x$
has no singularity. Altogether we get
\begin{equation}
 \delta F(T)=\D_V\D_W\left\{
 \frac{64G_3}{\pi}\,\mathcal J(T)
 -\frac{16G_3}{\pi\bp}
 \left[
 -1+\frac{(\nu-\bv)^2}{2\nu}S_\nu
    -\frac{(\nu+\bv)^2}{2\nu}S_{-\nu}
 \right]\right\},
 \label{eq:exactOTO}
\end{equation}
where $S_{\pm\nu}$ are evaluated at $z=\pi/2+\ii T$ and
\begin{equation}
 \mathcal J(T)
 =\int_{-\infty}^{0}\frac{\dd x}{p(x)}
 \left[
 \frac{y_+}{(1-y_+)^2}+\frac{y_-}{(1-y_-)^2}
 \right],
 \qquad
 y_\pm=-e^{4x\mp2T}.
 \label{eq:J}
\end{equation}
Equation \eqref{eq:exactOTO} is exact at this order and valid for all
$T$. On the same trajectory, the Lerch functions have the real integral
representation
\begin{equation}
 S_\alpha\left(\frac{\pi}{2}+\ii T\right)
 =-\frac14\int_0^1 t^{-\alpha/2}
 \left[
 \frac{e^{-2T}}{1+t e^{-2T}}
 +\frac{e^{2T}}{1+t e^{2T}}
 \right]\dd t,
 \qquad \alpha<2,
 \label{eq:Sa-real-integral}
\end{equation}
obtained by writing $(2m-\alpha)^{-1}=\tfrac12\int_0^1t^{m-\alpha/2-1}\dd t$
under the sum. For $0<\nu<2$ this makes \eqref{eq:exactOTO}
straightforward to evaluate numerically.

\subsection*{Late-time growth}
\label{sec:growth}

The late-time behaviour of \eqref{eq:exactOTO} is controlled by the
two poles of $C_n$ in the right half-plane. Since
$p=p_0e^{2x}+O(e^{4x})$ near the horizon,
$\cI_n=1/[2p_0(n-1)]+\ldots$, and for $\nu\neq1$ the residues are
\begin{equation}
 \Res_{n=1}C_n=\frac{4G_3}{\pi p_0},
 \qquad
 \Res_{n=\nu}C_n=-\frac{G_3(\nu-\bv)^2}{\pi\bp\,\nu^3}.
 \label{eq:residues}
\end{equation}
The corresponding exponentials can be extracted directly from the
integral representations. In the $y_-$ term of $\mathcal J$ the
integrand is concentrated around $x=-T/2$, which lies deep in the
near-horizon region when $T$ is large. Setting $x=-T/2+w$ and using
$p\simeq p_0e^{2x}$, we get
\begin{equation}
 \mathcal J(T)
 =-\frac{e^T}{p_0}\int_{-\infty}^{\infty}
       \frac{e^{2w}}{(1+e^{4w})^2}\,\dd w+O(1)
 =-\frac{\pi}{8p_0}\,e^T+O(1).
 \label{eq:J-asymptotic}
\end{equation}
Similarly, rescaling $t=e^{-2T}s$ in \eqref{eq:Sa-real-integral}
gives
\begin{equation}
 S_\nu\left(\frac{\pi}{2}+\ii T\right)
 =-\frac{e^{\nu T}}4\int_0^\infty\frac{s^{-\nu/2}}{1+s}\,\dd s+O(1)
 =-\frac{\pi}{4\sin(\pi\nu/2)}\,e^{\nu T}+O(1),
 \qquad 0<\nu<2,
 \label{eq:Snu-asymptotic}
\end{equation}
while $S_{-\nu}$ stays bounded. Substituting these into
\eqref{eq:exactOTO} gives the result
\begin{equation}
 \delta F(T)=\D_V\D_W\left[
 -\frac{8G_3}{p_0}\,e^T
 +\frac{2G_3(\nu-\bv)^2}{\bp\,\nu\sin(\pi\nu/2)}\,e^{\nu T}
 \right]+O(g),
 \label{eq:OTOC-asymptotics-derived}
\end{equation}
with a remainder that is bounded as $T\to\infty$.

The asymptotic coefficient in \eqref{eq:Snu-asymptotic} extends to
higher noninteger values of $\nu$. For $0<z<\pi$, taking the
difference of the two cosine series in \eqref{eq:Sa} gives
\begin{equation}
 S_\nu(z)-S_{-\nu}(z)
 =2\nu\sum_{m=1}^{\infty}\frac{\cos(2mz)}{4m^2-\nu^2}
 =\frac{1}{\nu}
  -\frac{\pi}{2}\frac{\cos[\nu(z-\pi/2)]}{\sin(\pi\nu/2)} .
 \label{eq:Snu-continuation}
\end{equation}
Continuing to $z=\pi/2+\ii T$ turns the cosine into $\cosh(\nu T)$.
Meanwhile, \eqref{eq:Sa-real-integral} gives
$S_{-\nu}(\pi/2+\ii T)\to-1/(2\nu)$ as $T\to\infty$ for every
fixed $\nu>0$. Thus the coefficient of $e^{\nu T}$ in
\eqref{eq:Snu-asymptotic} holds on every noninteger interval in
\cref{tab:cosmic-otoc-signs}. At integer $\nu\geq2$, the full
Euclidean Hessian has a physical zero mode, and we exclude these
values from the Gaussian calculation.

\subsection{The time-ordered correlator and the squared commutator}
\label{sec:norm}

The time-ordered correlator $C(T)$, the OTOC at the edges of its strip
of analyticity, and the squared commutator are all built from the
single operator $A=yW(T)Vy$ introduced in \cref{sec:otoc-def}. In a
Hilbert space, this gives an exact relation among the three. The
tree-level correlators obey the corresponding relation,
$\delta C(T)=\Re\,\delta F(T+\ii\pi/2)$. This determines the late-time
growth of $C(T)$, and it implies that the squared commutator vanishes
at first order in $g$.

\paragraph{Operator identities:}
Let us suppose for the moment that $\rho$ is a positive density matrix
and that $V$ and $W$ are Hermitian. Then
\begin{equation}
 A=yW(T)Vy,\qquad A^\dagger=yVW(T)y.
 \label{eq:Adef}
\end{equation}
Using the cyclicity of the trace and $W(T\pm\ii\pi/2)=y^{\pm1}W(T)y^{\mp1}$,
we find
\begin{equation}
 \frac{\Tr(AA^\dagger)}{N_VN_W}=C(T),\qquad
 \frac{\Tr(A^2)}{N_VN_W}=F(T+\ii\pi/2),\qquad
 \frac{\Tr\bigl((A^\dagger)^2\bigr)}{N_VN_W}=F(T-\ii\pi/2).
 \label{eq:A-traces}
\end{equation}
Since $\Tr\bigl((A^\dagger)^2\bigr)=[\Tr(A^2)]^*$, the OTOC takes
complex conjugate values on the two edges of the strip. The
Cauchy--Schwarz inequality for the Hilbert--Schmidt inner product,
$|\Tr(A^2)|\leq\Tr(AA^\dagger)$, then gives
$|F(T\pm\ii\pi/2)|\leq C(T)$, the inequality used in the argument
of~\cite{MSS}.

\paragraph{The squared commutator:}
The squared commutator places the whole
thermal factor on one side. But here, we distribute it symmetrically
and define
\begin{equation}
 \mathcal Q(T)\equiv
 -\frac{\Tr\bigl(\rho^{1/2}[W(T),V]\,\rho^{1/2}[W(T),V]\bigr)}{N_VN_W}.
 \label{eq:Qdef}
\end{equation}
Writing out the two commutators gives four terms,
\begin{equation}
 \begin{aligned}
 N_VN_W\,\mathcal Q(T)
 ={}&\Tr\bigl(\rho^{1/2}W(T)V\rho^{1/2}VW(T)\bigr)
   +\Tr\bigl(\rho^{1/2}VW(T)\rho^{1/2}W(T)V\bigr)\\
 &-\Tr\bigl(\rho^{1/2}W(T)V\rho^{1/2}W(T)V\bigr)
   -\Tr\bigl(\rho^{1/2}VW(T)\rho^{1/2}VW(T)\bigr)\\
 ={}&\Tr(AA^\dagger)+\Tr(A^\dagger A)
   -\Tr(A^2)-\Tr\bigl((A^\dagger)^2\bigr).
 \end{aligned}
 \label{eq:Q-expansion}
\end{equation}
The first two terms are time ordered, and they are equal by cyclicity.
The last two are out of time order. By \eqref{eq:A-traces},
\begin{equation}
 \mathcal Q(T)
 =2C(T)-F(T+\ii\pi/2)-F(T-\ii\pi/2)
 =2C(T)-2\Re F(T+\ii\pi/2).
 \label{eq:commnorm}
\end{equation}
The four terms in \eqref{eq:Q-expansion} also combine into
$\Tr\bigl[(A-A^\dagger)^\dagger(A-A^\dagger)\bigr]$, with
$A-A^\dagger=y[W(T),V]y$. Thus $N_VN_W\,\mathcal Q$ is the squared
Hilbert--Schmidt norm of $y[W(T),V]y$, and $\mathcal Q$ is nonnegative
when $\rho$ is positive.

One might have expected $\mathcal Q$ to be $2[1-\Re F(T)]$, as in the
usual heuristic relation between the squared commutator and the OTOC.
That relation assumes that the time-ordered terms factorize. In
\eqref{eq:commnorm} the time-ordered terms give $C(T)$, which need not
be close to one.

\paragraph{Tree-level relation:}
Writing $C(T)$ and
$F(T+\ii\pi/2)$ in terms of the propagator $G_\xi$ of \eqref{eq:Green},
we show in \cref{app:TOC} that they differ only in how a cut of $G_\xi$
is approached. 
\begin{equation}
 \delta C(T)=\Re\,\delta F(T+\ii\pi/2).
 \label{eq:tree-norm-equality}
\end{equation}
By \eqref{eq:commnorm} it follows that
\begin{equation}
 \mathcal Q(T)=O(g^2),\qquad T>0.
 \label{eq:commnorm-tree-cancellation}
\end{equation}
This is the cancellation one expects in a Hilbert space. The operators
$V$ and $W$ interact only through gravity, so $[W(T),V]$ is of order
$g$, and \eqref{eq:Qdef} is quadratic in it. Here, however, it follows
from the Gaussian expansion of \cref{sec:gaussian}, which does not
assume that $\rho$ is positive.

\paragraph{Late-time growth of $C(T)$:}
The growth of $C(T)$ now follows from
\eqref{eq:OTOC-asymptotics-derived}.  The Weyl
term acquires the phase $e^{\ii\pi\nu/2}$, and we get
\begin{equation}
 \delta C_\nu(T)
 =\cos\frac{\pi\nu}{2}\,\delta F_\nu(T)
 =\frac{2G_3\D_V\D_W(\nu-\bv)^2}{\bp\,\nu}\,
   \cot\frac{\pi\nu}{2}\,e^{\nu T}.
 \label{eq:edgecheck}
\end{equation}
The coefficient is positive in both regimes: $\bp$ and
$\cot(\pi\nu/2)$ are both positive in the exterior ($0<\nu<1$) and
both negative in the cosmic patches considered here ($1<\nu<2$).

In the exterior, the time-ordered correction grows more slowly than
the leading $e^T$ term in the OTOC. In a cosmic patch, the Weyl mode
gives the leading $e^{\nu T}$ growth in both correlators. The deviation of $C(T)$ from
its factorized value is therefore of the same order as the growing
part of the OTOC, and the factorization assumption of the chaos
bound~\cite{MSS} fails.


\section{Explicit backgrounds}
\label{sec:explicit}

In this section we shall evaluate the coefficients in
\eqref{eq:OTOC-asymptotics-derived} on the individual backgrounds.
They are determined by the three numbers $\bp$, $\bv$ and $p_0$. The
first two follow directly from \eqref{eq:pk}, while $p_0$ depends on
the normalization $x(\rb)=0$ of the tortoise coordinate and requires
one integration. For nonrotating BTZ, the full kernel $C_n$ is
elementary, and the Schwarzian limit can be checked explicitly.

\subsection{Rotating BTZ}
\label{sec:explicit-rotBTZ}

We treat the two patches \eqref{eq:patches} together. In each, $\rh$
is the horizon that bounds the patch, and we denote the other horizon
by $r_o$, so that $(\rh,r_o)=(r_+,r_-)$ in the exterior and
$(\rh,r_o)=(r_-,r_+)$ in the inner-horizon patch. The
ratio $\gamma=r_o/\rh$ then satisfies $0\leq\gamma<1$ in the exterior
and $\gamma>1$ in the inner-horizon patch, and
$\varsigma=\sgn(1-\gamma)$. From \eqref{eq:pk} we get
\begin{equation}
 \bp=\rh\,\frac{(1-\nu^2)(1-\gamma^2\nu^2)}{(1-\gamma^2)\nu^2},
 \qquad
 \bv=-\nu\,\frac{1+\gamma^2-2\gamma^2\nu^2}{1-\gamma^2},
 \qquad
 \nu-\bv=\frac{2\nu(1-\gamma^2\nu^2)}{1-\gamma^2}.
 \label{eq:BTZpv}
\end{equation}
The coefficient $p_0$ depends on the normalization $x(\rb)=0$ of the
tortoise coordinate. Integrating
$\dd x/\dd r=\rh/(r^2-\rh^2)-(r_o^2/\rh)/(r^2-r_o^2)$ gives
\begin{equation}
 e^{2x(r)}
 =\frac{|r-\rh|}{r+\rh}\,\frac1{q_h}
  \left[\frac{|r-r_o|}{r+r_o}\,\frac1{q_o}\right]^{-\gamma},
 \qquad
 q_h=\frac{|\rb-\rh|}{\rb+\rh},\quad
 q_o=\frac{|\rb-r_o|}{\rb+r_o},
 \label{eq:BTZtortoise}
\end{equation}
and expanding near $r=\rh$ we find
\begin{equation}
 p_0=4\varsigma\rh\,q_h\left(\frac{q_*}{q_o}\right)^{\gamma},
 \qquad
 q_*=\frac{|\rh-r_o|}{\rh+r_o}.
 \label{eq:BTZp0}
\end{equation}
The two growing terms in \eqref{eq:OTOC-asymptotics-derived} are
therefore
\begin{equation}
 \delta F_1=-\frac{8G_3\D_V\D_W}{p_0}\,e^T,
 \qquad
 \delta F_\nu
 =g\D_V\D_W\,
 \frac{8\nu^3(1-\gamma^2\nu^2)}
 {(1-\gamma^2)(1-\nu^2)\sin(\pi\nu/2)}\,e^{\nu T},
 \label{eq:rotBTZcoefficient}
\end{equation}
and $\delta C_\nu=\cos(\pi\nu/2)\,\delta F_\nu$. In the exterior
$p_0>0$ and $\nu<1$, so the thermal term is negative and dominates,
and the boundary term is positive. In the inner-horizon patch
$p_0<0$, and for $1<\nu<2$ the three factors $1-\gamma^2\nu^2$,
$1-\gamma^2$ and $1-\nu^2$ are all negative. The boundary term is then
negative, and it dominates.

It is instructive to express $\nu$ in terms of the boundary data. The
angular potential \eqref{eq:clock} is positive in the exterior and
negative in the inner-horizon patch for $j>0$. Defining
\begin{equation}
 c_K=\frac{K\ell-\sqrt{(K\ell)^2-4}}2,
 \qquad K\ell>2,
 \label{eq:cK}
\end{equation}
one finds from \eqref{eq:K} and \eqref{eq:clock} that
$c_K^2=(1-\nu^2)/(1-\gamma^2\nu^2)$ and $\Om=\varsigma\gamma c_K$, and
hence
\begin{equation}
 \nu^2=\frac{1-c_K^2}{1-\Om^2},
 \qquad
 \gamma=\frac{|\Om|}{c_K}.
 \label{eq:sources}
\end{equation}
These are the rotating CBC saddles of~\cite{Liouville}, written in our
sign convention for the angular potential. The exterior saddle exists
for $|\Om|<c_K$ and the inner-horizon saddle for $c_K<|\Om|<1$, so
generic boundary data select one or the other. At fixed $K$ the
exponent $\nu$ grows with $|\Om|$ and crosses $1$ at $|\Om|=c_K$. On
that locus $\gamma\to1$ as well. The wall is then in the near-horizon
region of a near-extremal black hole, and the two poles of $C_n$
merge. Our formulas assume separated poles and cannot be used there.

\paragraph{Comparison with an asymptotic boundary:- }
It is useful to compare these results with those for rotating BTZ with
an asymptotic
boundary~\cite{ReynoldsRoss,Stikonas,Poojary,JKY,MezeiSarosi,Craps}.
As $\rb\to\infty$ in the exterior, $\nu\to0$ and the
Weyl term disappears. The
thermal term remains. In Killing time it grows as $e^{\kappa t}$, with
$\kappa=2\pi T_H$. The Hawking temperature is related to the two
chiral temperatures $T_\pm=(r_+\pm r_-)/(2\pi\ell^2)$ by
$2/T_H=1/T_++1/T_-$. This is the growth of the uniform component of the
shock wave on the horizon~\cite{ReynoldsRoss}. The separate rates
$2\pi T_\pm$ found for operators localized in angle do not appear in
\eqref{eq:mainOTOC}. They come from the angular profile of the
shock, to which the angle-independent sector is blind. Recovering them
at a finite wall requires the nonzero angular harmonics, which we leave
to future work.

\subsection{Nonrotating BTZ}
\label{sec:explicit-nonrot}

Here $\gamma=0$,
\begin{equation}
 \bp=\rh\,\frac{1-\nu^2}{\nu^2},\qquad
 \bv=-\nu,\qquad
 p_0=4\rh\,\frac{1-\nu}{1+\nu},
 \label{eq:nonrotBTZdata}
\end{equation}
and
\begin{align}
 F(T)&=1+g\D_V\D_W\left[
 -\frac{2(1+\nu)}{1-\nu}\,e^T
 +\frac{8\nu^3}{(1-\nu^2)\sin(\pi\nu/2)}\,e^{\nu T}
 \right]+O(g)+O(g^2),
 \label{eq:nonrotBTZ}\\
 C(T)&=1+g\D_V\D_W\,
 \frac{8\nu^3}{1-\nu^2}\cot\frac{\pi\nu}{2}\,e^{\nu T}
 +O(g)+O(g^2).
 \label{eq:nonrotBTZ-C}
\end{align}
For instance, a wall at $\rb=2r_+$ gives
$F=1+g\D_V\D_W[-6e^T+\tfrac{4\sqrt2}{3}e^{T/2}]+\ldots$ and
$C=1+\tfrac43g\D_V\D_We^{T/2}+\ldots$.

In this case the full kernel is elementary. With $q=(1-\nu)/(1+\nu)$
and $p_0=4\rh q$,
\begin{equation}
 \frac1{p(x)}
 =\frac{e^{-2x}}{p_0}-\frac1{2\rh}+\frac{p_0e^{2x}}{16\rh^2},
 \qquad
 \cI_n
 =\frac1{2p_0(n-1)}-\frac1{4\rh n}+\frac{p_0}{32\rh^2(n+1)},
 \label{eq:nonrot-radial-integral}
\end{equation}
and \eqref{eq:CnGeneral} collapses to
\begin{equation}
 C_n=\frac{2g}{\pi}\,
 \frac{2\nu n^2+n-\nu}{n^2(n^2-1)(n-\nu)},
 \qquad n\geq2.
 \label{eq:nonrotkernel}
\end{equation}
The pole at $n=-\nu$ is absent because $\bv=-\nu$. As $\nu\to0$ this
becomes the Schwarzian propagator
\eqref{eq:dirichlet-Schwarzian-limit}. Using
$\sum_{m\geq1}\cos(2mz)/(4m^2-1)=\tfrac12-\tfrac{\pi}{4}\sin z$ for
$0\leq z\leq\pi$ in \eqref{eq:antipodal-exchange} we recover the
familiar result
\begin{equation}
 F(T)=1+g\D_V\D_W\left[\frac8\pi-4\cosh T\right]+O(g^2).
 \label{eq:Schwarzian-OTOC}
\end{equation}

\subsection{De Sitter}
\label{sec:explicit-dS}

For the Kerr-de Sitter $f(r)$ in \eqref{eq:dS}, let $\zeta=a/\rh$. The expressions are
obtained from those of BTZ by $\gamma^2\to-\zeta^2$:
\begin{equation}
 \bp=\rh\,\frac{(1-\nu^2)(1+\zeta^2\nu^2)}{(1+\zeta^2)\nu^2},
 \qquad
 \bv=-\nu\,\frac{1-\zeta^2+2\zeta^2\nu^2}{1+\zeta^2},
 \qquad
 \nu-\bv=\frac{2\nu(1+\zeta^2\nu^2)}{1+\zeta^2}.
 \label{eq:dSpv}
\end{equation}
The tortoise coordinate is
\begin{equation}
 e^{2x(r)}
 =\frac{\rh-r}{\rh+r}\,\frac{\rh+\rb}{\rh-\rb}\,
 \exp\left[\frac{2a}{\rh}
 \left(\arctan\frac{r}{a}-\arctan\frac{\rb}{a}\right)\right],
 \label{eq:dStortoise}
\end{equation}
so that
\begin{equation}
 p_0=-4\rh\,\frac{\rh-\rb}{\rh+\rb}\,
 \exp\left[\frac{2a}{\rh}
 \left(\arctan\frac{\rb}{a}-\arctan\frac{\rh}{a}\right)\right],
 \label{eq:dSp0}
\end{equation}
with the exponential replaced by $1$ for $a=0$. The boundary term is
\begin{equation}
 \delta F_\nu
 =g\D_V\D_W\,
 \frac{8\nu^3(1+\zeta^2\nu^2)}
 {(1+\zeta^2)(1-\nu^2)\sin(\pi\nu/2)}\,e^{\nu T},
 \label{eq:dSrotcoef}
\end{equation}
which is negative for $1<\nu<2$, while the thermal term
$-8G_3\D_V\D_We^T/p_0$ is positive. Without rotation,
\begin{align}
 F(T)&=1+g\D_V\D_W\left[
 \frac{2(\nu+1)}{\nu-1}\,e^T
 -\frac{8\nu^3}{(\nu^2-1)\sin(\pi\nu/2)}\,e^{\nu T}
 \right]+O(g)+O(g^2),
 \label{eq:nonrotdS}\\
 C(T)&=1-g\D_V\D_W\,
 \frac{8\nu^3}{\nu^2-1}\cot\frac{\pi\nu}{2}\,e^{\nu T}
 +O(g)+O(g^2).
 \label{eq:nonrotdS-C}
\end{align}
These are the nonrotating BTZ formulas continued to $\nu>1$, and the
kernel \eqref{eq:nonrotkernel} holds with $q<0$. There
is no geometric reason for $\nu<2$. The upper limit only
marks the first resonance $\nu=2$ of the Gaussian integral
(\cref{sec:gaussian}). To go beyond it, one needs to continue the $\alpha$ in \eqref{eq:Sa-real-integral}. The result is presented in table \ref{tab:cosmic-otoc-signs}.

\subsection{Relation to finite-coupling SYK}
\label{sec:syk-comparison}

In the nonrotating BTZ exterior, the boundary reparametrization
propagator \eqref{eq:nonrotkernel} reduces to the Schwarzian propagator
as $\nu=\rh/\rb\to0$. In the SYK model, the Schwarzian regime is
reached at strong coupling, $\beta\mathcal J\gg1$, where $\beta$ is
the inverse temperature and $\mathcal J$ is the interaction scale.
A relation between the fluctuation equations persists away from these
limits.

To see this, write $r=\rh\coth y$ in the BTZ exterior. The wall lies
at $\tanh y_b=\nu$, the horizon is at $y\to\infty$, and the background
conformal factor is $e^{2\rho_0}=\ell^2/\sinh^2y$. Since the dilaton
potential $U(\Phi)=2\Phi/\ell^2$ is linear, the equation for the bulk
conformal-factor fluctuation $\hr_n$ in \eqref{eq:linearized-radial-eom}
decouples and takes the form
\begin{equation}
 \left[-\partial_y^2+\frac{2}{\sinh^2y}\right]\hr_n
 =-n^2\hr_n .
 \label{eq:btz-syk-radial}
\end{equation}
The complex shift $y=s+\ii\pi/2$ turns the potential into
$-2/\cosh^2s$. This is precisely the P\"oschl--Teller potential in the
retarded ladder equation of large-$N$, large-$q$
SYK~\cite{MaldacenaStanford}. There, $s=\pi v(t_1-t_2)/\beta$ is a
rescaled relative time, and the eigenvalue is
$-(\beta\lambda_L/2\pi v)^2$. Normalizability on the real line selects
the bound state $\psi(s)\propto1/\cosh s$, with eigenvalue $-1$.
The resulting Lyapunov exponent and the thermal saddle relation are
\begin{equation}
 \lambda_L^{\rm SYK}=\frac{2\pi v}{\beta},
 \qquad
 \beta\mathcal J=\frac{\pi v}{\cos(\pi v/2)},
 \qquad 0<v<1 .
\end{equation}
In BTZ, the physical problem is posed on the real radial interval
$y_b\leq y<\infty$, with regularity at the horizon and matching to
the fluctuating wall at $y_b$.

The four-point functions distinguish the roles of $v$ and $\nu$.
In large-$q$ SYK, the poles at $\pm v$ contribute to the
out-of-time-ordered four-point function and cancel in the time-ordered
one~\cite{ChoiMezeiSarosi}. Varying $\beta\mathcal J$ therefore moves
the leading scrambling pole. For the conformal wall, the thermal
contribution proportional to $e^T$ persists at finite radius, and the
Weyl mode adds a term proportional to $e^{\nu T}$. Its contribution
survives in the time-ordered correlator, with
\[
 \delta C_\nu(T)=\cos(\pi\nu/2)\,\delta F_\nu(T)
\]
for the insertions used in \eqref{eq:edgecheck}. This ordering
dependence has a direct geometric origin. The wall displacement
already grows at rate $\nu$ in the classical fixed-$K$ equation
\eqref{eq:lorentzian-weyl-frequency}, and matter probes inherit that
growth. In the exterior it is subleading to the thermal OTOC exponent
$2\pi/\tb$.

\section{Retarded correlators}
\label{sec:retarded}

In this section, we shall compute retarded correlators, which isolate the
growth of the boundary mode. The Weyl mode is an inverted oscillator
with rate $\nu$. The reparametrization mode responds with both rates.

\subsection*{Causal prescription}
\label{sec:causal}

We define
\begin{equation}
 G^R_{AB}(T)=-\ii\theta(T)\langle[A(T),B(0)]\rangle,
 \qquad
 G^R_{AB}(\omega)=\int_0^\infty e^{\ii\omega T}G^R_{AB}(T)\,\dd T.
 \label{eq:GRdef}
\end{equation}
Since the responses of interest grow exponentially, the transform
exists only for $\Im\omega$ larger than every growth rate. Then the
inverse must be taken along a line above all singularities,
\begin{equation}
 G^R(T)=\int_{\mathbb R+\ii c}\frac{\dd\omega}{2\pi}\,
        e^{-\ii\omega T}G^R(\omega).
 \label{eq:Bromwich}
\end{equation}
For $T<0$ the contour closes in the upper half plane and gives zero.
For $T>0$ it closes below and picks up the unstable poles. A contour
along the real axis would miss them, and it would not solve the
initial value problem. The retarded functions are obtained from the
Euclidean kernels by continuing from integer $n$ to $n=s=-\ii\omega$,
with $\Re s>\max(1,\nu)$. The same continuation takes the regular
Euclidean solution $e^{nx-\ii n\theta}$ to $e^{-\ii\omega(T+x)}$. In
the exterior $x=\kappa r_*$ up to a constant, and in a cosmic patch
$x=-\kappa r_*$. In both cases this is a wave falling through the
future horizon.

\input{figures/contours_retarded.tex}

\subsection{The Weyl mode}
\label{sec:retarded-weyl}

In real time \eqref{eq:WeylAction} becomes
\begin{equation}
 S_\sigma
 =\frac{m_\sigma}{2}\int\left(\dot\sigma^2+\nu^2\sigma^2\right)\dd T,
 \qquad
 m_\sigma=-\frac{\rb^2}{4G_3\bp},
 \label{eq:Weyl-real-time}
\end{equation}
which is an inverted oscillator, with a kinetic term of the wrong sign
in the exterior. The solution of $\ddot\sigma=\nu^2\sigma$ with
initial data $\sigma(0)$ and $\Pi_\sigma(0)=m_\sigma\dot\sigma(0)$,
together with $[\sigma(0),\Pi_\sigma(0)]=\ii$, gives
\begin{equation}
 G^R_{\sigma\sigma}(T)
 =-\frac{\theta(T)}{m_\sigma\nu}\,\sinh(\nu T)
 =\frac{4G_3\bp}{\rb^2\nu}\,\theta(T)\sinh(\nu T),
 \qquad
 G^R_{\sigma\sigma}(\omega)
 =-\frac{4G_3\bp}{\rb^2(\omega^2+\nu^2)}.
 \label{eq:weyl-retarded-derived}
\end{equation}
This agrees with the continuation of the Euclidean propagator,
$G^R_{\sigma\sigma}(\omega)=-2\pi\langle\sigma_s\sigma_{-s}\rangle$,
with \eqref{eq:sigprop}. The response is positive in the exterior and
negative in a cosmic patch.

\subsection{The reparametrization mode}
\label{sec:retarded-xi}

The retarded function of $\xi$ is obtained by the same continuation,
with one difference in sign. It grows at both rates, $e^T$ and
$e^{\nu T}$.

In Lorentzian signature the disk angle $\theta=\tau+\xi(\tau)$ becomes
$\ii[T+\varepsilon(T)]$, where $\varepsilon$ is the Lorentzian
reparametrization. Setting $\tau=\ii T$ we get
\begin{equation}
 \xi(\ii T)=\ii\,\varepsilon(T).
 \label{eq:vectorwick}
\end{equation}
Each factor of $\xi$ therefore carries an extra $\ii$. Compared with
the rule for the scalar $\sigma$ \eqref{eq:weyl-retarded-derived}. The retarded function of $\varepsilon$ has the opposite sign,
\begin{equation}
 G^R_{\varepsilon\varepsilon}(\omega)=+2\pi C_s,
 \qquad
 s=-\ii\omega,\quad \Re s>\max(1,\nu),
 \label{eq:wickSigns}
\end{equation}
where $C_s$ is \eqref{eq:CnGeneral} continued to $n=s$.

In terms of $s$, \eqref{eq:GRdef} is a Laplace transform, and it can
be inverted in closed form. Setting $x=-T'/2$ in \eqref{eq:DtN} turns
the radial integral into a Laplace transform,
$\cI_s=\tfrac12\int_0^\infty e^{-sT'}\dd T'/p(-T'/2)$, and the rational
part of $C_s$ inverts term by term. The result is
\begin{equation}
 \begin{aligned}
 G^R_{\varepsilon\varepsilon}(T)
 =\theta(T)\bigg\{&
 8G_3\int_0^T\frac{\dd T'}{p(-T'/2)}\\
 &-\frac{4G_3}{\bp}\left[
 \left(1-\frac{\bv^2}{\nu^2}\right)T
 +\left(1+\frac{\bv^2}{\nu^2}\right)\frac{\sinh(\nu T)}{\nu}
 -\frac{2\bv}{\nu^2}\bigl(\cosh(\nu T)-1\bigr)
 \right]\bigg\}.
 \end{aligned}
 \label{eq:epsR}
\end{equation}
The first term involves $p(x)$ only for $-T/2<x<0$, so a response at
time $T$ probes the geometry down to $x=-T/2$. At short times the terms
linear in $T$ cancel, and
\begin{equation}
 G^R_{\varepsilon\varepsilon}(T)=\frac{4G_3}{\rb}\,T^2+O(T^3),
 \label{eq:epsR-short}
\end{equation}
in agreement with the large-$n$ behaviour \eqref{eq:UV}. At late times
the near-horizon form $p\simeq p_0e^{2x}$ gives
\begin{equation}
 G^R_{\varepsilon\varepsilon}(T)
 =\frac{8G_3}{p_0}\,e^T
  -\frac{2G_3(\nu-\bv)^2}{\bp\,\nu^3}\,e^{\nu T}+O(T).
 \label{eq:epsRasym}
\end{equation}
Both rates are present in $\varepsilon$. 

\subsection{Retarded commutator of bilocals}
\label{sec:retarded-bilocal}

A Hermitian observable with support in a finite time interval is
\begin{equation}
 \mathbb B_O(T;d)
 =\frac{\{O(T+d/2),O(T-d/2)\}}
       {\langle\{O(d/2),O(-d/2)\}\rangle_0}-1,
 \qquad d>0,
 \label{eq:hermitianbilocal}
\end{equation}
where the subscript $0$ denotes the correlator without gravitational
corrections \footnote{For \(d>0\), the denominator is \(2\cos(\pi\Delta)[4\sinh^2(d/2)]^{-\Delta}\). We take \(\cos(\pi\Delta)\ne0\), which includes the free-scalar probes with \(\Delta=1\).}. Here $d$ is measured in units of $b$. 
Continuing \eqref{eq:vertex} to real time, its first-order variation
is $\delta\mathbb B_O=\D_O\,\cD_d(\partial_T)\,\varepsilon(T)$, with
\begin{equation}
 \cD_d(\lambda)
 =2\left[
 \lambda\cosh\frac{\lambda d}{2}
 -\coth\frac d2\,\sinh\frac{\lambda d}{2}
 \right],
 \label{eq:Da}
\end{equation}
that is,
$\delta\mathbb B_O=\D_O[\dot\varepsilon(T+\tfrac d2)
+\dot\varepsilon(T-\tfrac d2)
-\coth\tfrac d2\,(\varepsilon(T+\tfrac d2)-\varepsilon(T-\tfrac d2))]$.
For two bilocals in different sectors whose supports do not overlap,
$T>(d_V+d_W)/2$, we get
\begin{equation}
 G^R_{\mathbb B_V\mathbb B_W}(T)
 \equiv-\ii\langle[\mathbb B_V(T;d_V),\mathbb B_W(0;d_W)]\rangle
 =-\D_V\D_W\,\cD_{d_V}(\partial_T)\,\cD_{d_W}(\partial_T)\,
 G^R_{\varepsilon\varepsilon}(T)+O(g^2).
 \label{eq:bilocalR}
\end{equation}
The minus sign arises because $\cD_d$ is odd, and derivatives with
respect to the position of the second bilocal act as $-\partial_T$.

The vertex annihilates the three gauge modes,
$\cD_d(0)=\cD_d(\pm1)=0$. Applied to \eqref{eq:epsR} it therefore
removes the $e^{T}$ term, and what grows is
\begin{equation}
 G^R_{\mathbb B_V\mathbb B_W}(T)
 \simeq\D_V\D_W\,
 \frac{2G_3(\nu-\bv)^2}{\bp\,\nu^3}\,
 \cD_{d_V}(\nu)\,\cD_{d_W}(\nu)\,e^{\nu T}.
 \label{eq:bilocalRgrow}
\end{equation}
Since $\cD_d(\nu)/[2\sinh(\nu d/2)]=\nu\coth(\nu d/2)-\coth(d/2)$ is
an increasing function of $\nu$ that vanishes at $\nu=1$, the product
$\cD_{d_V}(\nu)\cD_{d_W}(\nu)$ is positive for $\nu\neq1$. The
retarded commutator therefore has the sign of $\bp$, like
\eqref{eq:weyl-retarded-derived}: it is positive in the exterior and
negative in a cosmic patch. Thus a commutator of matter operators
gives direct access to the boundary mode, without the thermal
background that accompanies it in the OTOC. The retarded two-point
function of a single operator and the fully retarded four-point
function are discussed in \cref{app:fullyR}.

Let us now discuss what this result means. The retarded correlator measures how a perturbation in the $W$ sector
affects the $V$ bilocal at later times. At fixed boundary insertion
times, $\mathbb B_V$ probes the corresponding angles on the disk.
These angles respond to the motion of the wall because the boundary
reparametrization mixes with the Weyl mode. The bilocal vertex removes
the thermal $e^T$ contribution, so the late-time response inherits the
$e^{\nu T}$ growth of the classical wall displacement at fixed $K$.
The same exponential survives in the time-ordered correlator $C(T)$
(\cref{sec:norm}). Together, these two orderings identify the wall
instability as the origin of the second exponent in the OTOC.

\section{Discussion}
\label{sec:discussion}

We have computed the leading gravitational contribution to the
four-point function of operators in
three-dimensional gravity with conformal boundary conditions.  Everything follows from the
propagator \eqref{eq:CnGeneral} of the boundary reparametrization,
whose two poles produce the two growth rates in \eqref{eq:mainOTOC}.
Our main conclusions are the following.
\begin{enumerate}
 \item The OTOC grows at two rates, $2\pi/\tb$ and $2\pi\nu/\tb$ with
 $\nu=\rh/\rb$. The first is the thermal growth of the Schwarzian
 theory, and the second is the growth rate of the linearized Weyl
 mode.
 \item Outside a BTZ black hole, the thermal term dominates, with a
 coefficient that depends on the position of the wall. In the
 inner-horizon patch of BTZ and in the de~Sitter patch $\nu>1$, 
 the Weyl term dominates.
 \item The time-ordered correlator $C(T)$ grows at the rate
 $2\pi\nu/\tb$, with a coefficient of the same order, so for $\nu>1$
 the chaos bound cannot be applied. This is due to the instability of the conformal wall at fixed $K$.  The squared commutator vanishes at
 this order.
\end{enumerate}
We end with some discussions and some future directions.

\paragraph{The origin of the second exponent:-}
The rate $\nu$ is set by the boundary condition and the saddle. It is
already present in the linearized motion of the wall,
\eqref{eq:lorentzian-weyl-frequency}. It appears in $C(T)$ and in the retarded commutator as well as
in the OTOC. In the exterior this term is subleading, and the dominant
growth is the standard one, with a coefficient that depends on the
position of the wall through $p_0$.

\paragraph{Signs:-}
The signs in \cref{tab:signs} are determined by the type of patch and
not by the sign of the cosmological constant. The inner-horizon patch
of BTZ behaves like de~Sitter. In particular, the thermal term changes
sign when the horizon surrounds the wall. Its coefficient is
$-8G_3/p_0$ and $p_0$ is negative in the cosmic patches. In a
shock-wave description we expect this to correspond to a time delay in
the exterior and an advance in the cosmic patches. It would be
interesting to relate this to the time advance found for de~Sitter horizons in~\cite{Aalsma} and to the
recent analyses of~\cite{CSTY,HZ}.

Our formulas assume that the two exponents are distinct. As $\nu\to1$, as the wall approaches a nonextremal horizon, the two
coefficients diverge with opposite signs, but the tree-level OTOC has a
finite limit. At late times its growing part becomes
$-(4T+8)\,g\D_V\D_W\,e^T$, from either side and independently of the
rotation. At integer $\nu\geq2$ the Weyl mode has a zero mode, and the
Gaussian approximation itself breaks down. The extremal limit
$\kappa\to0$, and the near-extremal locus of
\cref{sec:explicit-rotBTZ} where $\nu\to1$ and $\gamma\to1$ together,
cannot be reached by substitution. These cases require a separate
analysis.

Several simplifications could be relaxed. First, we kept only the
sector that does not depend on the angle. Operators localized in
angle ~\cite{Poojary,Craps} are not included.
We discuss their relation to our result at the end of
\cref{sec:explicit-rotBTZ}. Second, our probes couple only to the reduced two-dimensional metric.
For the $s$-wave of a minimally coupled three-dimensional scalar,
the kinetic term also contains the factor $\Phi/r_h$ in
\eqref{eq:reduced-scalar-matter}. When $|r_b-r_h|\ll r_h$, the
background ratio $\Phi_0/r_h$ is close to unity throughout the patch,
so our probe action approximates free propagation on the stationary
geometry. Dilaton fluctuations also generate a vertex proportional to
$(\delta\Phi/r_h)(\nabla\chi)^2$, which can contribute to the connected
four-point function at order $G_3$. We leave its effect on the
coupling of the three-dimensional scalar to the thermal and Weyl modes
to future work. Third, our
operators are neutral and are defined with respect to the conformal
clock. Operators that carry angular momentum, or that are dressed to
the proper time of the wall, have different vertices and may couple to
the Weyl mode directly. We leave these extensions to future work.

Finally, our correlators are defined by a Gaussian expansion about an
unstable saddle. Since the Weyl mode has a nonlinear description as a
Liouville theory~\cite{NearExtremal,Liouville}, it will be interesting
to see whether the nonlinear dynamics limits the growth found here. It
is also not clear whether the correlators admit a Hilbert space
interpretation, in which $C(T)$ and $\mathcal Q(T)$ have the positivity
properties derived in \cref{sec:norm}. 

\acknowledgments

I thank D.~Anninos for discussions. I am grateful to the organizers of
the Celestial Holography Meeting at Harvard for hospitality during the
completion of this work. Claude (Anthropic) and Codex (OpenAI) were
used to carry out independent numerical and symbolic checks of some of
the calculations. They were also used to improve the clarity of the text, the
references and the figures. I take full responsibility for the content
of this paper. This work is supported by the CNS Sparks grant. I
would also like to thank OpenAI for offering
the Pro academic subscription.

\appendix

\section{The radial problem and the Hessian}
\label{app:hessian}

In this appendix, we derive several results used in \cref{sec:action}. We show integral \eqref{eq:constraint},
the boundary relation \eqref{eq:DtN} and the quadratic action
\eqref{eq:quadratic-completed-square}.\\

With $e^{2\rho_0}U=p'$ and $e^{2\rho_0}U'=2k'$, the first equation in
\eqref{eq:linearized-radial-eom} gives
\begin{equation}
 E_n'=(n^2+k')\hP_n-k\hP_n'+p'\hr_n-p\hr_n',
 \label{eq:Eprime}
\end{equation}
and differentiating once more and using both equations, we get
\begin{equation}
 (\partial_x^2-n^2)E_n
 =\left(k''-2kk'-\tfrac12p\,e^{2\rho_0}U''\right)\hP_n
 +\left(p''-2kp'-2pk'\right)\hr_n.
 \label{eq:first-integral-derivation}
\end{equation}
Both brackets vanish, as can be seen by differentiating $p'=2kp$ and
$k'=\tfrac12e^{2\rho_0}U'(\Phi_0)$. Hence $E_n=c_ne^{nx}+d_ne^{-nx}$,
and regularity at the centre requires $d_n=0$. Acting with
$\tfrac12(\partial_x+n)$ on $E_n$ and using \eqref{eq:Eprime} gives
the linearization of $\partial_z^2\Phi-2\partial_z\rho\,\partial_z\Phi$,
which shows that $E_n$ is the linearized constraint.\\

Eliminating $\hr_n$ between \eqref{eq:constraint} and the first
equation in \eqref{eq:linearized-radial-eom} gives
\begin{equation}
 \partial_x\left[
 e^{2nx}\partial_x\left(\frac{e^{-nx}\hP_n}{p}\right)\right]
 =2c_ne^{2nx}\partial_x\left(\frac1p\right).
 \label{eq:factorized-radial-eom}
\end{equation}
Integrating from the centre we get
\begin{equation}
 e^{2nx}\partial_x\left(\frac{e^{-nx}\hP_n}{p}\right)
 =2c_n\left[\frac{e^{2nx}}{p(x)}
 -2n\int_{-\infty}^{x}\frac{e^{2ny}}{p(y)}\,\dd y\right].
 \label{eq:first-radial-integration}
\end{equation}
A constant of integration on the right hand side would add to $\hP_n$
a term proportional to $p\,e^{-nx}\sim e^{(2-n)x}$, which is not
smooth at the centre for $n\geq2$. Substituting back into
\eqref{eq:constraint} gives
\begin{equation}
 \hr_n(x)=\frac{n+k(x)}{p(x)}\,\hP_n(x)
 +c_ne^{nx}\left[\frac1{p(x)}
 -4ne^{-2nx}\int_{-\infty}^{x}\frac{e^{2ny}}{p(y)}\,\dd y\right],
 \label{eq:Y}
\end{equation}
which at $x=0$ is \eqref{eq:DtN}. The remaining constant of
integration and $c_n$ are fixed by the two boundary values.

\paragraph{Quadratic action:}
Let $\hP_b$ and $\hr_b$ denote the values at the wall, and $\hP_b'$
and $\hr_b'$ the radial derivatives there. From \eqref{eq:constraint},
\eqref{eq:Eprime} and \eqref{eq:DtN} we get
\begin{equation}
 \begin{pmatrix}\hr_b'\\ \hP_b'\end{pmatrix}
 =
 \begin{pmatrix}
 \dfrac{n^2+k_b'-\kb^2}{\bp}+\dfrac{(n+\kb)^2}{\bp^2D_n}
 &
 \kb-\dfrac{n+\kb}{\bp D_n}\\[2mm]
 \kb-\dfrac{n+\kb}{\bp D_n}
 &
 \bp+\dfrac1{D_n}
 \end{pmatrix}
 \begin{pmatrix}\hP_b\\ \hr_b\end{pmatrix}.
 \label{eq:DNmatrix}
\end{equation}
 Inserting it into
\eqref{eq:dirichlet-quadratic-response} gives (for each pair of modes
$\pm n$), a local part
\begin{equation}
 \frac{n^2+k_b'-\kb^2}{\bp}\,\hP_{-n}\hP_n
 +\kb(\hP_{-n}\hr_n+\hr_{-n}\hP_n)
 +\bp\,\hr_{-n}\hr_n
 \label{eq:local-dirichlet-form}
\end{equation}
and a nonlocal part
\begin{equation}
 \frac1{D_n}
 \left(\hr_{-n}-\frac{n+\kb}{\bp}\hP_{-n}\right)
 \left(\hr_{n}-\frac{n+\kb}{\bp}\hP_{n}\right)
 =\frac{(a_n\sigma_{-n}+\ii n\xi_{-n})
        (a_n\sigma_n-\ii n\xi_n)}{D_n},
 \label{eq:nonlocal-dirichlet-form}
\end{equation}
all evaluated at the wall and multiplied by $-2\pi/(8G_3)$.\\

The conversion between $\tau$ and $\theta$ enters as follows. On the
background, the first variation of $I_D$ is
\begin{equation}
 \delta I_D=-\frac1{4G_3}\int_0^{2\pi}\dd\theta\,
          \left(\kb\,\delta\Phi_b+\bp\,\delta\rho_b\right).
 \label{eq:dirichlet-first-variation}
\end{equation}
From \eqref{eq:exact-boundary-matching} the boundary values at a
fixed value of $\theta$ are, to second order,
\begin{equation}
 \begin{aligned}
  \Phi_b(\theta)/\rb
   &=1+\sigma+\tfrac12\sigma^2-\xi\sigma'+..,\\
  \rho_b(\theta)-\rho_0(0)
   &=\sigma-\xi'-\xi\sigma'+\xi\xi''+\tfrac12(\xi')^2+..,
 \end{aligned}
 \label{eq:second-order-boundary-data}
\end{equation}
where the terms with an undifferentiated $\xi$ come from
$\sigma(\tau(\theta))=\sigma(\theta)-\xi\sigma'(\theta)+\ldots$ and its
analogue for $\xi'$. Inserting the quadratic terms into
\eqref{eq:dirichlet-first-variation} and adding the quadratic part of
\eqref{eq:boundary-legendre-term} gives
\begin{equation}
 \frac1{8G_3}\int_0^{2\pi}\dd\tau\,
 \left[2(\bp+\rb\kb)\,\xi\sigma'
       +\bp(\xi')^2+(2\bp+\rb\kb)\,\sigma^2\right].
 \label{eq:boundary-quadratic-addition}
\end{equation}
With $\hP_n=\rb\sigma_n$ and $\hr_n=\sigma_n-\ii n\xi_n$, the $\xi\xi$
and $\sigma\xi$ terms of \eqref{eq:local-dirichlet-form} cancel
against \eqref{eq:boundary-quadratic-addition}. The coefficient of
$\sigma_{-n}\sigma_n$ that survives is
\begin{equation}
 -\frac{\rb^2}{\bp}(n^2+k_b'-\kb^2)+\bp-\rb\kb
 =-\frac{\rb^2}{\bp}(n^2-\nu^2),
 \label{eq:sigma-sigma-coefficient}
\end{equation}
with $\nu^2$ as in \eqref{eq:nuGeneral}. Together with
\eqref{eq:nonlocal-dirichlet-form} this is
\eqref{eq:quadratic-completed-square}. 

\section{Exact boundary action without rotation}
\label{app:nonrot}

For $j=0$ the potential is linear, and the dilaton can be integrated
out exactly. It fixes $R=\mp2/\ell^2$. The disk is then a region of
the hyperbolic plane (BTZ) or of the sphere (de~Sitter),
\begin{equation}
 \dd s_2^2=\ell^2\left(\dd\varrho^2+S(\varrho)^2\dd\theta_{\rm p}^2\right),
 \qquad
 (S,C)=(\sinh,\cosh)\ \text{or}\ (\sin,\cos),
 \label{eq:nonrot-disk-metric}
\end{equation}
with $r=\rh C(\varrho)$ on the stationary solution. Here
$\theta_{\rm p}$ is the polar angle, which we distinguish from the
uniformizing angle $\theta$ of \eqref{eq:boundary-map}. We write
$c_b=C(\varrho_b)=\rb/\rh=1/\nu$ and $s_b=S(\varrho_b)$ for the values
at the wall. The bulk term of the action vanishes, and $I_K$ becomes a
functional of the boundary curve $(\varrho(\tau),\theta_{\rm
p}(\tau))$. Using $\Phi_b=\dd\ell/(b\,\dd\tau)$ we get
\begin{equation}
 I_K=\frac{\ell}{8G_3b}\int_0^{2\pi}
       \left[K\ell\,v^2-2v\,\bigl(C(\varrho)\,\theta_{\rm p}'-\psi'\bigr)
       \right]\dd\tau ,
 \qquad
 v^2=\varrho'^2+S(\varrho)^2\theta_{\rm p}'^2,
 \label{eq:exact-curve-action}
\end{equation}
where $\psi=\arctan[\varrho'/(S\theta_{\rm p}')]$ is the angle between
the tangent and the direction of $\partial_{\theta_{\rm p}}$. The
second term is the integrated geodesic curvature, and the first is the
boundary term of \eqref{eq:action}. Expanding about the circle,
$\varrho=\varrho_b+\eta$ and $\theta_{\rm p}=\tau+\lambda$, and using
$K\ell=(2c_b^2-1)/(c_bs_b)$ and $\ell/b=\rh c_b/s_b$, we find in both
cases
\begin{equation}
 I^{(2)}=\frac{\rh c_b}{8G_3s_b}\int_0^{2\pi}\dd\tau
 \left[-\frac{s_b}{c_b}(\lambda')^2+\frac{\eta^2}{c_bs_b}
       -\frac{1+c_b^2}{c_bs_b}(\eta')^2-2\lambda''\eta'\right].
 \label{eq:oldaction}
\end{equation}
The variables $(\eta,\lambda)$ are related to $(\sigma,\xi)$ by
\begin{equation}
 \sigma_n=\ii n\lambda_n+\frac{c_b}{s_b}\,\eta_n,\qquad
 \xi_n=\lambda_n+\frac{\ii}{s_b}\,\eta_n,
 \qquad n>0.
 \label{eq:eta-lambda-map}
\end{equation}
The first relation is $\sigma=\delta\log v$. The second expresses the matter couples to the uniformizing angle $\theta$. The conformal map that takes the deformed
region to the round disk shifts the boundary angle by the Hilbert
transform of the radial displacement $\delta x=\eta/s_b$. With
\begin{equation}
 \bp=\rh(c_b^2-1),\qquad \kb=c_b,\qquad
 D_n=-\frac{(n+c_b)^2}{\rh(n^2-1)(c_b^2-1)},
 \label{eq:nonrot-app-data}
\end{equation}
this change of variables takes \eqref{eq:oldaction} into
\eqref{eq:Hessian}, and its inverse gives \eqref{eq:nonrotkernel}.

\section{Comparison with the Schwarzian-Liouville action}
\label{app:GMS}

In this appendix, we give the details of the comparison with section~5
of~\cite{NearExtremal}. In our normalization, their near-boundary
variables are
\begin{equation}
 \frac{\Phi_b(u_E)}{\ell}=\frac{\Phi_r(u_E)}{\epsilon},
 \qquad
 \omega(u_E)=\log\Phi_r(u_E),
 \qquad
 \ell K=2+\epsilon^2\mu,
 \label{eq:GMS-boundary-scaling}
\end{equation}
with $\epsilon$ a dimensionless cutoff. The coefficient $\mu$ is
called $\kappa_2$ in~\cite{NearExtremal}, and their two-dimensional
Newton constant is $G_2=G_3/(2\pi\ell)$ when the dilaton is measured
in units of $\ell$. If $F(u_E)$ denotes the boundary value of the
Poincar\'e time, the leading boundary action is
\begin{equation}
 I_{\mathrm{asym}}
 =-\frac{\ell}{8G_3}\int_0^{\tb}\dd u_E\,
 \left[2\{F,u_E\}+(\partial_{u_E}\omega)^2
 -2\partial_{u_E}^2\omega-\mu e^{2\omega}\right],
 \label{eq:GMS-asymptotic-action}
\end{equation}
where $\{F,u\}=F'''/F'-\tfrac32(F''/F')^2$. In terms of $\Gamma$,
defined by $\partial_{u_E}\Gamma=e^\omega$, and
$\{\Gamma,u_E\}=\partial_{u_E}^2\omega
-\tfrac12(\partial_{u_E}\omega)^2$, it is
\begin{equation}
 I_{\mathrm{asym}}
 =-\frac{\ell}{4G_3}\int_0^{\tb}\dd u_E
 \left[\{F,u_E\}-\{\Gamma,u_E\}
 -\frac{\mu}{2}(\partial_{u_E}\Gamma)^2\right],
 \label{eq:GMS-two-Schwarzians}
\end{equation}
which is eq.~(5.20) of~\cite{NearExtremal}. Our boundary condition
$\Phi_b=\rb e^\sigma$ translates into
\begin{equation}
 \omega(u_E)=\log\frac{\epsilon\rb}{\ell}+\sigma(u_E/b),
 \label{eq:GMS-Weyl-dictionary}
\end{equation}
so the fluctuation of $\omega$ is our Weyl mode, and
$F=\tan[(\tau+\xi(\tau))/2]$.\\

The two descriptions overlap for nonrotating BTZ with
$\nu=\rh/\rb\to0$. The extrinsic curvature is
\begin{equation}
 \ell K=\frac{2-\nu^2}{\sqrt{1-\nu^2}}=2+\frac{\nu^4}{4}+O(\nu^6),
 \qquad
 b=\frac{\ell}{\rh}\sqrt{1-\nu^2},
 \label{eq:GMS-static-K-limit}
\end{equation}
and choosing $\epsilon=\nu$ gives $\Phi_r=\rh/\ell$ and
$\mu=\nu^2/4+O(\nu^4)$. The stationary solution sits at
$\mu\to0$, consistent with the equation of motion
$\partial_{u_E}^2\omega+\mu e^{2\omega}=0$. Expanding \eqref{eq:GMS-asymptotic-action}
at $\mu=0$ and
\begin{equation}
 \int_0^{2\pi}\left\{\tan\frac{\tau+\xi}2,\tau\right\}\dd\tau
 =\text{const}
 -\frac12\int_0^{2\pi}\left[(\xi'')^2-(\xi')^2\right]\dd\tau
 +O(\xi^3),
 \label{eq:Schwarzian-quadratic}
\end{equation}
$\{F,u_E\}=b^{-2}\{F,\tau\}$,  $\partial_{u_E}\omega=b^{-1}\sigma'$,
we obtain
\begin{equation}
 I_{\mathrm{asym}}^{(2)}
 =\frac{\rh}{8G_3}\int_0^{2\pi}\dd\tau\,
 \left[(\xi'')^2-(\xi')^2-(\sigma')^2\right].
 \label{eq:GMS-quadratic-match}
\end{equation}
In our convention, for nonrotating BTZ,
\begin{equation}
 a_n=-\frac{\nu(n+\nu)}{1-\nu^2},
 \qquad
 D_n=-\frac{(1+n\nu)^2}{\rh(1-\nu^2)(n^2-1)},
 \qquad
 H_0(n)=-\frac{\rh}{1-\nu^2}(n^2-\nu^2),
 \label{eq:GMS-nonrotating-coefficients}
\end{equation}
so that $a_n\to0$, and \eqref{eq:Hessian} tends to
\eqref{eq:GMS-Hessian-limit} as $\nu\to0$ at fixed $n$. This is the
Fourier transform of \eqref{eq:GMS-quadratic-match}. The propagators
become
\begin{equation}
 \langle\sigma_n\sigma_{-n}\rangle\to-\frac{2G_3}{\pi\rh\,n^2},
 \qquad
 \langle\xi_n\xi_{-n}\rangle\to\frac{2G_3}{\pi\rh\,n^2(n^2-1)},
 \label{eq:GMS-propagator-limits}
\end{equation}
and the mixed propagator vanishes. 

\section{The time-ordered correlator at tree level}
\label{app:TOC}

In this appendix, we derive the relation \eqref{eq:tree-norm-equality}
between the time-ordered correlator and the OTOC at the edge of
strip, without assuming that $\rho$ is positive. Both
correlators can be written in terms of the propagator $G_\xi$ of
\eqref{eq:Green}.

Since $\langle\xi(\tau_1)\xi(\tau_2)\rangle=G_\xi(\tau_1-\tau_2)$, the
derivatives in the vertex \eqref{eq:vertex} give
$\langle\xi'(\tau_1)\xi'(\tau_2)\rangle=-G_\xi''(\tau_1-\tau_2)$. For
antipodal pairs the cotangents vanish, and using the periodicity of
$G_\xi$ we get
\begin{equation}
 \delta F(z)
 =-2\D_V\D_W\left[G_\xi''(-z)+G_\xi''(\pi-z)\right],
 \label{eq:F-from-G}
\end{equation}
which is \eqref{eq:antipodal-exchange} written in terms of $G_\xi$.
Since $G_\xi$ has cuts only at $\Re\tau\in2\pi\mathbb Z$,
\eqref{eq:F-from-G} is analytic in the strip $0<\Re z<\pi$.
The point $F(T+\ii\pi/2)$ corresponds to $z=\ii T$, on the edge of the
strip. Approaching it from inside, $z=0^++\ii T$, we get
\begin{equation}
 \delta F(T+\ii\pi/2)
 =-2\D_V\D_W
 \left[G_\xi''(0^--\ii T)+G_\xi''(\pi-\ii T)\right].
 \label{eq:F-edge-kernel}
\end{equation}
For $C(T)$ we use the configuration \eqref{eq:Cpositions}. In the $W$
vertex the cotangent is $\cot(\epsilon-\pi/2)=-\tan\epsilon$, which
vanishes as $\epsilon\to0$, so both vertices reduce to the sum of their
two $\xi'$ terms. The four differences between a $V$ and a $W$
insertion tend to $0^\pm-\ii T$ and $\pi^\pm-\ii T$, and $G_\xi''$ is
continuous across $\Re\tau=\pi$. This gives
\begin{equation}
 \delta C(T)=-\D_V\D_W\left[
 G_\xi''(0^--\ii T)+G_\xi''(0^+-\ii T)+2G_\xi''(\pi-\ii T)\right].
 \label{eq:C-edge-kernel}
\end{equation}

The two results differ only at the cut at $\Re\tau=0$. $G_\xi$ is real
and even on the Euclidean circle,
$G_\xi''(0^--\ii T)=G_\xi''(0^++\ii T)=[G_\xi''(0^+-\ii T)]^*$. Hence, the
two values are complex conjugates. The same argument shows that
$G_\xi''(\pi-\ii T)$ is real. Comparing \eqref{eq:F-edge-kernel} with
\eqref{eq:C-edge-kernel}  gives \eqref{eq:tree-norm-equality}.

Computing $\mathcal Q$ at order $g^2$ requires going beyond the
quadratic action and the linear vertex, as well as a treatment of the
coincident limit in \eqref{eq:Cpositions}. We have not done this analysis.

\section{Other retarded correlators}
\label{app:fullyR}

\paragraph{Two-point function.}
Without gravitational corrections, the retarded function of a primary
of dimension $\D$ in the conformal frame is
\begin{equation}
 G^R_{OO,0}(T)
 =-\ii\theta(T)\lim_{\epsilon\to0^+}\left\{
 \left[4\sin^2\frac{\epsilon+\ii T}{2}\right]^{-\D}
 -\left[4\sin^2\frac{2\pi-\epsilon+\ii T}{2}\right]^{-\D}
 \right\}.
 \label{eq:matterR}
\end{equation}
For $T>0$ the two arguments approach the negative real axis from
opposite sides, with phases $e^{\mp\ii\pi\D}$. So away from $T=0$
\begin{equation}
 G^R_{OO,0}(T)
 =-2\sin(\pi\D)\,\theta(T)\,[4\sinh^2(T/2)]^{-\D}.
 \label{eq:matterR-explicit}
\end{equation}
For integer $\D$, as for the free scalar, this vanishes, and only the
contact terms in \eqref{eq:matterR} remain. The usual normalization in
terms of $u$ is restored by a factor $b^{-2\D}$. The gravitational
correction to this two-point function is a self-energy, and it is not
part of the exchange computed in the main text.

\paragraph{Fully retarded four-point function.}
The kernel \eqref{eq:master} determines every ordering of the four
operators. For $A=V(T_1)$, $B=W(T_2)$, $C=V(T_3)$ and $D=W(T_4)$ the
fully retarded function is
\begin{align}
 R^{(4)}_{1234}
 &=(-\ii)^3\theta_{12}\theta_{23}\theta_{34}\,
     \langle[[[A,B],C],D]\rangle
 \nonumber\\
 &=(-\ii)^3\theta_{12}\theta_{23}\theta_{34}
 \bigl(
 \mathscr W_{ABCD}-\mathscr W_{BACD}
 -\mathscr W_{CABD}+\mathscr W_{CBAD}
 \nonumber\\
 &\hspace{32mm}
 -\mathscr W_{DABC}+\mathscr W_{DBAC}
 +\mathscr W_{DCAB}-\mathscr W_{DCBA}
 \bigr),
 \label{eq:fullyR}
\end{align}
with $\theta_{ij}=\theta(T_i-T_j)$. Each connected Wightman function is
obtained from
\begin{equation}
 [4\sin^2(\tau_{13}/2)]^{-\D_V}\,
 [4\sin^2(\tau_{24}/2)]^{-\D_W}\,
 \mathcal M(13;24)
 \label{eq:Wword}
\end{equation}
by setting $\tau_i=\epsilon_i+\ii T_i$, with the $\epsilon_i$
decreasing from left to right in the corresponding word, and then
taking $\epsilon_i\to0$. The prefactors must be continued along with
$\mathcal M$, since their phases differ between the eight terms.
Together with \eqref{eq:Hfunction} this gives the four-time response
in terms of one-dimensional integrals. Its growth depends on how the
four times are taken apart.

\bibliographystyle{JHEP}
\bibliography{references}
\end{document}

%% file: figures/patches.tex
\begin{tikzpicture}[
  >=Latex, font=\small,
  hor/.style={black!70,line width=.6pt},
  extfill/.style={fill=blue!12},
  cosfill/.style={fill=red!12},
  extwall/.style={draw=blue!55!black,line width=1.1pt},
  coswall/.style={draw=red!65!black,line width=1.1pt},
  lab/.style={font=\scriptsize}]

 \begin{scope}[x=1.2cm,y=1.2cm]
  \node[font=\small\bfseries] at (0.2,4.75) {Rotating BTZ, $(t,r)$ plane};

  \fill[extfill] (0,0) -- (1,1) .. controls (0.6,0.4) and (0.6,-0.4) .. (1,-1) -- cycle;
  \fill[extfill] (0,0) -- (-1,1) .. controls (-0.6,0.4) and (-0.6,-0.4) .. (-1,-1) -- cycle;
  \fill[cosfill] (0,2) -- (1,3) .. controls (0.6,2.4) and (0.6,1.6) .. (1,1) -- cycle;
  \fill[cosfill] (0,2) -- (-1,3) .. controls (-0.6,2.4) and (-0.6,1.6) .. (-1,1) -- cycle;

  \draw[hor] (-1,-1) -- (1,1) (1,-1) -- (-1,1);
  \draw[hor] (-1,1) -- (1,3) (1,1) -- (-1,3);
  \draw[hor] (-1,3) -- (0,4) -- (1,3);
  \draw[hor] (-1,-1) -- (0,-2) -- (1,-1);

  \draw[line width=1pt] (1,-1) -- (1,1) (-1,-1) -- (-1,1);
  \draw[densely dashed] (1,1) -- (1,3) (-1,1) -- (-1,3);
  \node[lab,anchor=west] at (1.03,0) {$r=\infty$};
  \node[lab,anchor=west] at (1.03,2) {$r=0$};

  \draw[extwall] (1,-1) .. controls (0.6,-0.4) and (0.6,0.4) .. (1,1);
  \draw[extwall] (-1,-1) .. controls (-0.6,-0.4) and (-0.6,0.4) .. (-1,1);
  \draw[coswall] (1,1) .. controls (0.6,1.6) and (0.6,2.4) .. (1,3);
  \draw[coswall] (-1,1) .. controls (-0.6,1.6) and (-0.6,2.4) .. (-1,3);

  \foreach \sx in {1,-1}{
   \node[lab] at (0.36*\sx,0.70)  {$r_+$};
   \node[lab] at (0.36*\sx,1.30)  {$r_-$};
   \node[lab] at (0.36*\sx,3.30)  {$r_+$};
   \node[lab] at (0.36*\sx,-1.30) {$r_-$};
  }

  \fill (0,0) circle (1.4pt) node[lab,below=2pt] {$B_+$};
  \fill (0,2) circle (1.4pt) node[lab,above=2pt] {$B_-$};

  \node at (0.42,0) {I};      \node at (-0.42,0) {I};
  \node at (0.40,2) {III};    \node at (-0.40,2) {III};
  \node at (0,1.25) {II};
  \draw[->,black!60] (0,0.98) -- (0,0.58);
  \node at (0,2.85) {II$'$};
  \draw[->,black!60] (0,3.08) -- (0,3.48);
  \node at (0,-0.82) {II$'$};
  \draw[->,black!60] (0,-1.5) -- (0,-1.1);
 \end{scope}

 \begin{scope}[shift={(3.3cm,0.5cm)},x=1.5cm,y=0.68cm]
  \node[font=\small\bfseries] at (1.75,7.65) {Reduced theory};

  \fill[cosfill] (0.75,0) -- plot[domain=0.75:1,samples=40]
       (\x,{(\x*\x-4)*(\x*\x-1)/(\x*\x)}) -- cycle;
  \fill[extfill] (2,0) -- plot[domain=2:3,samples=60]
       (\x,{(\x*\x-4)*(\x*\x-1)/(\x*\x)}) -- (3,0) -- cycle;

  \draw[->] (0,0) -- (3.45,0) node[right] {$\Phi$};
  \draw[->] (0,-1.5) -- (0,6.9) node[above] {$f(\Phi)$};

  \draw[thick] plot[domain=0.62:3.2,samples=150,smooth]
       (\x,{(\x*\x-4)*(\x*\x-1)/(\x*\x)});

  \draw[coswall] (0.75,0) -- (0.75,2.674);
  \draw[extwall] (3,0) -- (3,4.444);

  \fill (1,0) circle (1.4pt) (2,0) circle (1.4pt);
  \node[lab,below,red!65!black]  at (0.75,0) {$r_b$};
  \node[lab,anchor=south west]   at (1,0)    {$r_-$};
  \node[lab,anchor=south east]   at (2,0)    {$r_+$};
  \node[lab,below,blue!55!black] at (3,0)    {$r_b$};

  \node[lab,red!65!black]  at (1.17,3.3) {$U<0$};
  \node[lab,blue!55!black] at (2.33,3.3) {$U>0$};
  \node[lab] at (1.5,-0.48) {$f<0$};

  \node[lab] at (0.45,-2.0) {III};
  \node[lab] at (1.5,-2.0)  {II, II$'$};
  \node[lab] at (2.6,-2.0)  {I};
 \end{scope}
\end{tikzpicture}

%% file: figures/contours_oto.tex
\begin{figure}[htbp]
\centering
\begin{tikzpicture}[>=Latex,scale=0.97]
 \begin{scope}[shift={(0,0)}]
  \draw[thick] (0,0) circle (1.45);
  \draw[->,blue!65!black,thick] (30:1.45) arc (30:70:1.45);
  \foreach \a/\txt/\col in {0/{V_1}/blue,90/{W_1}/red,180/{V_2}/blue,270/{W_2}/red}{
    \fill[\col!70!black] (\a:1.45) circle (2.7pt);
    \node at (\a:1.9) {$\txt$};
  }
  \node[align=center] at (0,0) {OTO\\$\tau=0,\frac\pi2,\pi,\frac{3\pi}2$};
  \node at (0,-2.35) {(a) Alternating Euclidean order};
 \end{scope}
 \begin{scope}[shift={(5.5,0)}]
  \draw[thick] (0,0) circle (1.45);
  \fill[blue!70!black] (0:1.45) circle (2.7pt);
  \fill[blue!70!black] (180:1.45) circle (2.7pt);
  \fill[red!70!black] (22:1.45) circle (2.7pt);
  \fill[red!70!black] (158:1.45) circle (2.7pt);
  \node[right] at (0:1.65) {$V_1$};
  \node[left] at (180:1.65) {$V_2$};
  \node[above right] at (28:1.5) {$W_1$};
  \node[above left] at (152:1.5) {$W_2$};
  \draw[->,thick,red!70!black] (22:1.7) arc (22:4:1.7);
  \draw[->,thick,red!70!black] (158:1.7) arc (158:176:1.7);
  \node[align=center] at (0,-0.15) {C(T)\\$V_1W_1W_2V_2$};
  \node at (0,-2.35) {(b) Noninterlaced regulator};
 \end{scope}
 \begin{scope}[shift={(-1.2,-4.0)}]
  \draw[->,gray] (-0.25,0.4)--(7.5,0.4) node[right] {$\Re z_{\rm L}$};
  \foreach \y/\lab in {0/0,-0.8/{-\pi/2},-1.6/{-\pi},-2.4/{-3\pi/2}}{
    \node[left] at (-0.1,\y) {$\lab$};
  }
  \draw[->,blue!65!black,thick] (0,0)--(6.8,0);
  \draw[->,blue!65!black,thick] (6.8,0)--(6.8,-0.8);
  \draw[->,blue!65!black,thick] (6.8,-0.8)--(0,-0.8);
  \draw[->,blue!65!black,thick] (0,-0.8)--(0,-1.6);
  \draw[->,blue!65!black,thick] (0,-1.6)--(6.8,-1.6);
  \draw[->,blue!65!black,thick] (6.8,-1.6)--(6.8,-2.4);
  \draw[->,blue!65!black,thick] (6.8,-2.4)--(0,-2.4);
  \draw[->,blue!65!black,thick,dashed] (0,-2.4)--(0,-3.05);
  \node[below] at (0,-3.05) {KMS closure: $-2\pi\ii$};
  \fill[blue!70!black] (0,0) circle (3pt);
  \fill[blue!70!black] (0,-1.6) circle (3pt);
  \fill[red!70!black] (4.5,-0.8) circle (3pt);
  \fill[red!70!black] (4.5,-2.4) circle (3pt);
  \node[above right] at (0,0) {$V(0)$};
  \node[above right] at (0,-1.6) {$V(0)$};
  \node[above] at (4.5,-0.8) {$W(T)$};
  \node[above] at (4.5,-2.4) {$W(T)$};
  \node[below] at (6.8,-2.4) {$T_{\max}>T$};
  \node[align=center] at (4.4,-3.45) {(c) Four real-time legs,\\$z_{\rm L}=T-\ii\tau$};
 \end{scope}
\end{tikzpicture}
\caption{The operator contours.  (a) Equally spaced OTOC \(F(T)\)= $\langle VWVW\rangle $. The two $W$ insertions are continued
to real time $T$ while their Euclidean separations are kept fixed.
(b) The time-ordered correlator \(C(T)\)= $\langle VWWV\rangle $  (c) Real-time contour for \(F(T)\). Each vertical interval represents a thermal factor
$y$.}
\label{fig:OTOcontour}
\end{figure}
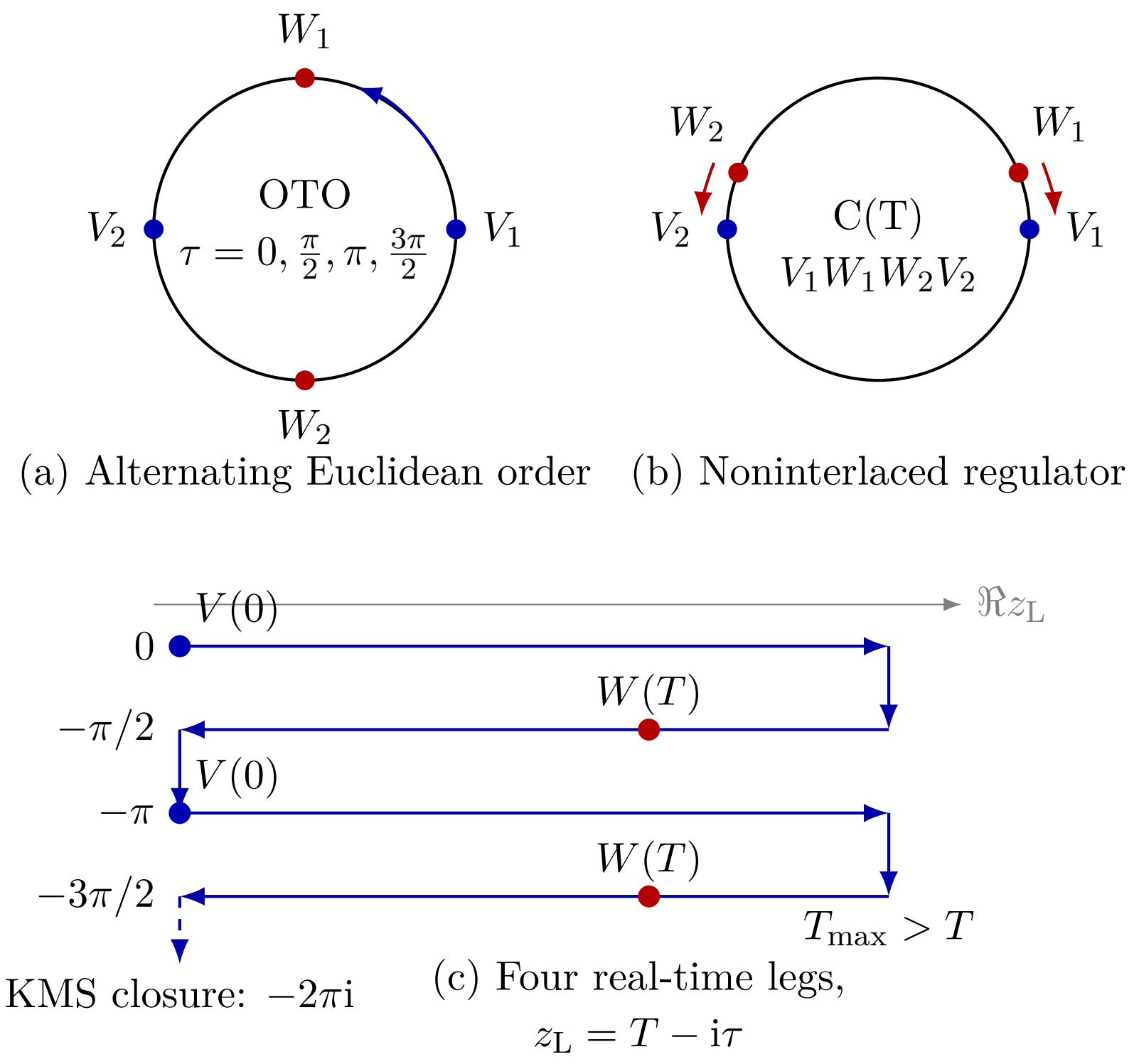

%% file: figures/contours_retarded.tex
\begin{figure}[ht]
\centering
\begin{tikzpicture}[>=Latex]
 \begin{scope}[shift={(8,0)}]
  \draw[->,gray] (-2,0)--(2,0) node[right] {$\Re\omega$};
  \draw[->,gray] (0,-1.8)--(0,2.25) node[above] {$\Im\omega$};
  \fill[red!70!black] (0,0.95) circle (2.7pt);
  \fill[red!70!black] (0,-0.95) circle (2.7pt);
  \node[right] at (0.1,0.95) {$+\ii\nu$};
  \node[right] at (0.1,-0.95) {$-\ii\nu$};
  \draw[->,blue!65!black,very thick] (-1.95,1.7)--(1.95,1.7);
  \node[above left,blue!65!black] at (0,1.7) {$\mathbb R+\ii c$};
  \node[above right,blue!65!black] at (0.3,1.7) {$c>\nu$};
  \draw[->,dashed,blue!65!black] (1.9,1.7)..controls(2.3,-1.7)and(1.8,-2.05)..(0.6,-2.05);
  \node at (0,-2.75) { Weyl response in frequency space};
 \end{scope}
\end{tikzpicture}
\caption{ The inverse
transform for the unstable Weyl mode runs above both poles. Closing
downward for $T>0$ gives $\sinh(\nu T)$; closing upward for $T<0$ gives
zero. For the auxiliary reparametrization kernel, the line is chosen
above $\max(1,\nu)$.}
\label{fig:retarded}
\end{figure}
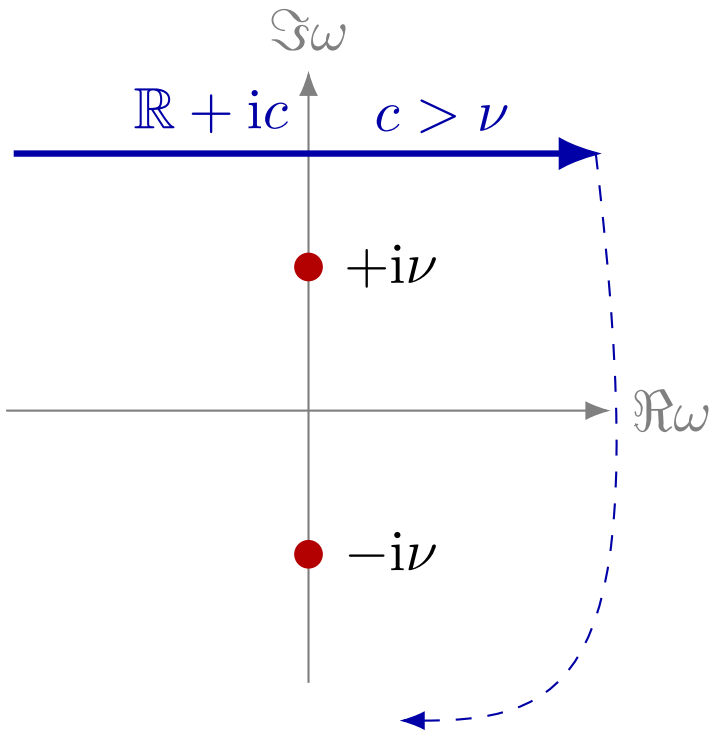

%% file: references.bib
@article{SS,
  author = {Shenker, S. H. and Stanford, D.},
  title = {Black holes and the butterfly effect},
  year = {2014},
  eprint = {1306.0622},
  journal = {JHEP},
  volume = {03},
  pages = {067},
}

@article{Poojary,
  author = {Poojary, R. R.},
  title = {{BTZ} dynamics and chaos},
  year = {2020},
  eprint = {1812.10073},
  journal = {JHEP},
  volume = {03},
  pages = {048},
}

@article{Craps,
  author = {Craps, B. and Khetrapal, S. and Rabideau, C.},
  title = {Chaos in {CFT} dual to rotating {BTZ}},
  year = {2021},
  eprint = {2107.13874},
  journal = {JHEP},
  volume = {11},
  pages = {105},
}

@article{Aalsma,
  author = {Aalsma, L. and Shiu, G.},
  title = {Chaos and complementarity in {de Sitter} space},
  year = {2020},
  eprint = {2002.01326},
  journal = {JHEP},
  volume = {05},
  pages = {152},
}

@article{CSTY,
  author = {Chen, Y. and Stanford, D. and Tang, H. and Yang, Z.},
  title = {Negative shocks versus static patch holography},
  year = {2026},
  eprint = {2607.14042},
}

@article{HZ,
  author = {Harlow, D. and Zhao, Y.},
  title = {Anti-scrambling and {Euclidean} folds from observer correlators in {de Sitter} space},
  year = {2026},
  eprint = {2607.14215},
}

@article{Observatories,
  author = {Anninos, D. and Galante, D. A. and Maneerat, C.},
  title = {Cosmological Observatories},
  year = {2024},
  eprint = {2402.04305},
  journal = {Class. Quant. Grav.},
  volume = {41},
  pages = {165009},
  doi = {10.1088/1361-6382/ad5824},
}

@article{NearExtremal,
  author = {Galante, D. A. and Maneerat, C. and Svesko, A.},
  title = {Conformal boundaries near extremal black holes},
  year = {2025},
  eprint = {2504.14003},
  journal = {Class. Quant. Grav.},
  volume = {42},
  pages = {195003},
  doi = {10.1088/1361-6382/ae0408},
}

@article{Liouville,
  author = {Allameh, K. and Shaghoulian, E.},
  title = {Timelike {Liouville} theory and {AdS}$_3$ gravity at finite cutoff},
  year = {2026},
  eprint = {2508.03236},
  journal = {JHEP},
  volume = {09},
  pages = {264},
  doi = {10.1007/JHEP09(2026)264},
}

@article{BMS,
  author = {Bousso, R. and Maloney, A. and Strominger, A.},
  title = {Conformal Vacua and Entropy in {de Sitter} Space},
  year = {2002},
  eprint = {hep-th/0112218},
  journal = {Phys. Rev. D},
  volume = {65},
  pages = {104039},
}

@article{MSS,
  author = {Maldacena, J. and Shenker, S. H. and Stanford, D.},
  title = {A bound on chaos},
  year = {2016},
  eprint = {1503.01409},
  journal = {JHEP},
  volume = {08},
  pages = {106},
}

@article{MSY,
  author        = {Maldacena, Juan and Stanford, Douglas
                   and Yang, Zhenbin},
  title         = {Conformal symmetry and its breaking in two
                   dimensional nearly {Anti-de-Sitter} space},
  journal       = {PTEP},
  volume        = {2016},
  number        = {12},
  pages         = {12C104},
  year          = {2016},
  doi           = {10.1093/ptep/ptw124},
  eprint        = {1606.01857},
  archivePrefix = {arXiv},
  primaryClass  = {hep-th}
}

@article{JTFiniteCutoff,
  author        = {Iliesiu, Luca V. and Kruthoff, Jorrit
                   and Turiaci, Gustavo J. and Verlinde, Herman},
  title         = {{JT} gravity at finite cutoff},
  journal       = {SciPost Phys.},
  volume        = {9},
  pages         = {023},
  year          = {2020},
  doi           = {10.21468/SciPostPhys.9.2.023},
  eprint        = {2004.07242},
  archivePrefix = {arXiv},
  primaryClass  = {hep-th}
}

@article{LarkinOvchinnikov,
  author  = "Larkin, A. I. and Ovchinnikov, Yu. N.",
  title   = "{Quasiclassical method in the theory of superconductivity}",
  journal = "Sov. Phys. JETP",
  volume  = "28",
  pages   = "1200",
  year    = "1969"
}

@misc{Kitaev2014,
  author = "Kitaev, Alexei",
  title  = "{Hidden correlations in the Hawking radiation and thermal noise}",
  howpublished = "Talk given at the Fundamental Physics Prize Symposium",
  month  = "10 November",
  year   = "2014"
}

@misc{Kitaev2015,
  author = "Kitaev, Alexei",
  title  = "{A simple model of quantum holography}",
  howpublished = "Talks at KITP, \url{http://online.kitp.ucsb.edu/online/entangled15/kitaev/} and \url{http://online.kitp.ucsb.edu/online/entangled15/kitaev2/}",
  month  = "7 April and 27 May",
  year   = "2015"
}

@article{HaydenPreskill,
  author        = "Hayden, Patrick and Preskill, John",
  title         = "{Black holes as mirrors: quantum information in random subsystems}",
  eprint        = "0708.4025",
  archivePrefix = "arXiv",
  primaryClass  = "hep-th",
  journal       = "JHEP",
  volume        = "09",
  pages         = "120",
  year          = "2007"
}

@article{SekinoSusskind,
  author        = "Sekino, Yasuhiro and Susskind, Leonard",
  title         = "{Fast scramblers}",
  eprint        = "0808.2096",
  archivePrefix = "arXiv",
  primaryClass  = "hep-th",
  journal       = "JHEP",
  volume        = "10",
  pages         = "065",
  year          = "2008"
}

@article{Lashkari,
  author        = "Lashkari, Nima and Stanford, Douglas and Hastings, Matthew and Osborne, Tobias and Hayden, Patrick",
  title         = "{Towards the fast scrambling conjecture}",
  eprint        = "1111.6580",
  archivePrefix = "arXiv",
  primaryClass  = "hep-th",
  journal       = "JHEP",
  volume        = "04",
  pages         = "022",
  year          = "2013"
}

@article{Hosur,
  author        = "Hosur, Pavan and Qi, Xiao-Liang and Roberts, Daniel A. and Yoshida, Beni",
  title         = "{Chaos in quantum channels}",
  eprint        = "1511.04021",
  archivePrefix = "arXiv",
  primaryClass  = "hep-th",
  journal       = "JHEP",
  volume        = "02",
  pages         = "004",
  year          = "2016"
}

@article{MaldacenaEternal,
  author        = "Maldacena, Juan Martin",
  title         = "{Eternal black holes in anti-de Sitter}",
  eprint        = "hep-th/0106112",
  archivePrefix = "arXiv",
  journal       = "JHEP",
  volume        = "04",
  pages         = "021",
  year          = "2003"
}

@article{SSmultiple,
  author        = "Shenker, Stephen H. and Stanford, Douglas",
  title         = "{Multiple shocks}",
  eprint        = "1312.3296",
  archivePrefix = "arXiv",
  primaryClass  = "hep-th",
  journal       = "JHEP",
  volume        = "12",
  pages         = "046",
  year          = "2014"
}

@article{RSSlocalized,
  author        = "Roberts, Daniel A. and Stanford, Douglas and Susskind, Leonard",
  title         = "{Localized shocks}",
  eprint        = "1409.8180",
  archivePrefix = "arXiv",
  primaryClass  = "hep-th",
  journal       = "JHEP",
  volume        = "03",
  pages         = "051",
  year          = "2015"
}

@article{SSstringy,
  author        = "Shenker, Stephen H. and Stanford, Douglas",
  title         = "{Stringy effects in scrambling}",
  eprint        = "1412.6087",
  archivePrefix = "arXiv",
  primaryClass  = "hep-th",
  journal       = "JHEP",
  volume        = "05",
  pages         = "132",
  year          = "2015"
}

@article{SachdevYe,
  author        = "Sachdev, Subir and Ye, Jinwu",
  title         = "{Gapless spin fluid ground state in a random, quantum Heisenberg magnet}",
  eprint        = "cond-mat/9212030",
  archivePrefix = "arXiv",
  journal       = "Phys. Rev. Lett.",
  volume        = "70",
  pages         = "3339",
  year          = "1993"
}

@article{MaldacenaStanford,
  author        = "Maldacena, Juan and Stanford, Douglas",
  title         = "{Remarks on the Sachdev-Ye-Kitaev model}",
  eprint        = "1604.07818",
  archivePrefix = "arXiv",
  primaryClass  = "hep-th",
  journal       = "Phys. Rev. D",
  volume        = "94",
  pages         = "106002",
  year          = "2016"
}

@article{KitaevSuh,
  author        = "Kitaev, Alexei and Suh, S. Josephine",
  title         = "{The soft mode in the Sachdev-Ye-Kitaev model and its gravity dual}",
  eprint        = "1711.08467",
  archivePrefix = "arXiv",
  primaryClass  = "hep-th",
  journal       = "JHEP",
  volume        = "05",
  pages         = "183",
  year          = "2018"
}

@article{RobertsStanford,
  author        = "Roberts, Daniel A. and Stanford, Douglas",
  title         = "{Two-dimensional conformal field theory and the butterfly effect}",
  eprint        = "1412.5123",
  archivePrefix = "arXiv",
  primaryClass  = "hep-th",
  journal       = "Phys. Rev. Lett.",
  volume        = "115",
  pages         = "131603",
  year          = "2015"
}

@article{JMV,
  author        = "Jackson, Steven and McGough, Lauren and Verlinde, Herman",
  title         = "{Conformal bootstrap, universality and gravitational scattering}",
  eprint        = "1412.5205",
  archivePrefix = "arXiv",
  primaryClass  = "hep-th",
  journal       = "Nucl. Phys. B",
  volume        = "901",
  pages         = "382",
  year          = "2015"
}

@article{FitzpatrickKaplan,
  author        = "Fitzpatrick, A. Liam and Kaplan, Jared",
  title         = "{A quantum correction to chaos}",
  eprint        = "1601.06164",
  archivePrefix = "arXiv",
  primaryClass  = "hep-th",
  journal       = "JHEP",
  volume        = "05",
  pages         = "070",
  year          = "2016"
}

@article{Perlmutter,
  author        = "Perlmutter, Eric",
  title         = "{Bounding the space of holographic CFTs with chaos}",
  eprint        = "1602.08272",
  archivePrefix = "arXiv",
  primaryClass  = "hep-th",
  journal       = "JHEP",
  volume        = "10",
  pages         = "069",
  year          = "2016"
}

@article{TuriaciVerlinde,
  author        = "Turiaci, Gustavo and Verlinde, Herman",
  title         = "{On CFT and quantum chaos}",
  eprint        = "1603.03020",
  archivePrefix = "arXiv",
  primaryClass  = "hep-th",
  journal       = "JHEP",
  volume        = "12",
  pages         = "110",
  year          = "2016"
}

@article{HaehlRozali,
  author        = "Haehl, Felix M. and Rozali, Moshe",
  title         = "{Effective field theory for chaotic CFTs}",
  eprint        = "1808.02898",
  archivePrefix = "arXiv",
  primaryClass  = "hep-th",
  journal       = "JHEP",
  volume        = "10",
  pages         = "118",
  year          = "2018"
}

@article{CotlerJensen,
  author        = "Cotler, Jordan and Jensen, Kristan",
  title         = "{A theory of reparameterizations for AdS$_3$ gravity}",
  eprint        = "1808.03263",
  archivePrefix = "arXiv",
  primaryClass  = "hep-th",
  journal       = "JHEP",
  volume        = "02",
  pages         = "079",
  year          = "2019"
}

@article{BlakeLeeLiu,
  author        = "Blake, Mike and Lee, Hyunseok and Liu, Hong",
  title         = "{A quantum hydrodynamical description for scrambling and many-body chaos}",
  eprint        = "1801.00010",
  archivePrefix = "arXiv",
  primaryClass  = "hep-th",
  journal       = "JHEP",
  volume        = "10",
  pages         = "127",
  year          = "2018"
}

@article{GSS,
  author        = "Grozdanov, Sa{\v{s}}o and Schalm, Koenraad and Scopelliti, Vincenzo",
  title         = "{Black hole scrambling from hydrodynamics}",
  eprint        = "1710.00921",
  archivePrefix = "arXiv",
  primaryClass  = "hep-th",
  journal       = "Phys. Rev. Lett.",
  volume        = "120",
  pages         = "231601",
  year          = "2018"
}

@article{DraytHooft,
  author  = "Dray, Tevian and 't Hooft, Gerard",
  title   = "{The gravitational shock wave of a massless particle}",
  journal = "Nucl. Phys. B",
  volume  = "253",
  pages   = "173",
  year    = "1985"
}

@article{tHooft,
  author  = "'t Hooft, Gerard",
  title   = "{The black hole interpretation of string theory}",
  journal = "Nucl. Phys. B",
  volume  = "335",
  pages   = "138",
  year    = "1990"
}

@article{KVV,
  author        = "Kiem, Youngjai and Verlinde, Herman L. and Verlinde, Erik P.",
  title         = "{Black hole horizons and complementarity}",
  eprint        = "hep-th/9502074",
  archivePrefix = "arXiv",
  journal       = "Phys. Rev. D",
  volume        = "52",
  pages         = "7053",
  year          = "1995"
}

@article{AlmheiriPolchinski,
  author        = "Almheiri, Ahmed and Polchinski, Joseph",
  title         = "{Models of AdS$_2$ backreaction and holography}",
  eprint        = "1402.6334",
  archivePrefix = "arXiv",
  primaryClass  = "hep-th",
  journal       = "JHEP",
  volume        = "11",
  pages         = "014",
  year          = "2015"
}

@article{Jensen,
  author        = "Jensen, Kristan",
  title         = "{Chaos in AdS$_2$ holography}",
  eprint        = "1605.06098",
  archivePrefix = "arXiv",
  primaryClass  = "hep-th",
  journal       = "Phys. Rev. Lett.",
  volume        = "117",
  pages         = "111601",
  year          = "2016"
}

@article{EMV,
  author        = "Engels{\"o}y, Julius and Mertens, Thomas G. and Verlinde, Herman",
  title         = "{An investigation of AdS$_2$ backreaction and holography}",
  eprint        = "1606.03438",
  archivePrefix = "arXiv",
  primaryClass  = "hep-th",
  journal       = "JHEP",
  volume        = "07",
  pages         = "139",
  year          = "2016"
}

@article{MTV,
  author        = "Mertens, Thomas G. and Turiaci, Gustavo J. and Verlinde, Herman L.",
  title         = "{Solving the Schwarzian via the conformal bootstrap}",
  eprint        = "1705.08408",
  archivePrefix = "arXiv",
  primaryClass  = "hep-th",
  journal       = "JHEP",
  volume        = "08",
  pages         = "136",
  year          = "2017"
}

@article{LMTV,
  author        = "Lam, Ho Tat and Mertens, Thomas G. and Turiaci, Gustavo J. and Verlinde, Herman",
  title         = "{Shockwave S-matrix from Schwarzian quantum mechanics}",
  eprint        = "1804.09834",
  archivePrefix = "arXiv",
  primaryClass  = "hep-th",
  journal       = "JHEP",
  volume        = "11",
  pages         = "182",
  year          = "2018"
}

@article{ReynoldsRoss,
  author        = "Reynolds, Alan P. and Ross, Simon F.",
  title         = "{Butterflies with rotation and charge}",
  eprint        = "1604.04099",
  archivePrefix = "arXiv",
  primaryClass  = "hep-th",
  journal       = "Class. Quant. Grav.",
  volume        = "33",
  pages         = "215008",
  year          = "2016"
}

@article{Stikonas,
  author        = "{\v{S}}tikonas, Andrius",
  title         = "{Scrambling time from local perturbations of the rotating BTZ black hole}",
  eprint        = "1810.06110",
  archivePrefix = "arXiv",
  primaryClass  = "hep-th",
  journal       = "JHEP",
  volume        = "02",
  pages         = "054",
  year          = "2019"
}

@article{JKY,
  author        = "Jahnke, Viktor and Kim, Keun-Young and Yoon, Junggi",
  title         = "{On the chaos bound in rotating black holes}",
  eprint        = "1903.09086",
  archivePrefix = "arXiv",
  primaryClass  = "hep-th",
  journal       = "JHEP",
  volume        = "05",
  pages         = "037",
  year          = "2019"
}

@article{MezeiSarosi,
  author        = "Mezei, M{\'a}rk and S{\'a}rosi, G{\'a}bor",
  title         = "{Chaos in the butterfly cone}",
  eprint        = "1908.03574",
  archivePrefix = "arXiv",
  primaryClass  = "hep-th",
  journal       = "JHEP",
  volume        = "01",
  pages         = "186",
  year          = "2020"
}

@article{SusskindAddendum,
  author        = "Susskind, Leonard",
  title         = "{Addendum to fast scramblers}",
  eprint        = "1101.6048",
  archivePrefix = "arXiv",
  primaryClass  = "hep-th",
  year          = "2011"
}

@article{AHH,
  author        = "Anninos, Dionysios and Hartnoll, Sean A. and Hofman, Diego M.",
  title         = "{Static patch solipsism: conformal symmetry of the de Sitter worldline}",
  eprint        = "1109.4942",
  archivePrefix = "arXiv",
  primaryClass  = "hep-th",
  journal       = "Class. Quant. Grav.",
  volume        = "29",
  pages         = "075002",
  year          = "2012"
}

@article{AGH,
  author        = "Anninos, Dionysios and Galante, Dami{\'a}n A. and Hofman, Diego M.",
  title         = "{De Sitter horizons \& holographic liquids}",
  eprint        = "1811.08153",
  archivePrefix = "arXiv",
  primaryClass  = "hep-th",
  journal       = "JHEP",
  volume        = "07",
  pages         = "038",
  year          = "2019"
}

@article{KolchmeyerLiu,
  author        = "Kolchmeyer, David K. and Liu, Hong",
  title         = "{Chaos and the emergence of the cosmological horizon}",
  eprint        = "2411.08090",
  archivePrefix = "arXiv",
  primaryClass  = "hep-th",
  year          = "2024"
}

@article{MNX,
  author        = "Milekhin, A. and Narovlansky, V. and Xu, J.",
  title         = "{Out-of-time-ordered correlators in de Sitter revisited}",
  eprint        = "2607.13137",
  archivePrefix = "arXiv",
  primaryClass  = "hep-th",
  year          = "2026"
}

@article{CuiKolchmeyer,
  author        = "Cui, W. and Kolchmeyer, D. K.",
  title         = "{A de Sitter anti-scrambling algebra}",
  eprint        = "2607.13665",
  archivePrefix = "arXiv",
  primaryClass  = "hep-th",
  year          = "2026"
}

@article{York72,
  author  = "York, Jr., James W.",
  title   = "{Role of conformal three-geometry in the dynamics of gravitation}",
  journal = "Phys. Rev. Lett.",
  volume  = "28",
  pages   = "1082",
  year    = "1972"
}

@article{York86,
  author  = "York, Jr., James W.",
  title   = "{Black hole thermodynamics and the Euclidean Einstein action}",
  journal = "Phys. Rev. D",
  volume  = "33",
  pages   = "2092",
  year    = "1986"
}

@article{BrownYork,
  author        = "Brown, J. David and York, Jr., James W.",
  title         = "{Quasilocal energy and conserved charges derived from the gravitational action}",
  eprint        = "gr-qc/9209012",
  archivePrefix = "arXiv",
  journal       = "Phys. Rev. D",
  volume        = "47",
  pages         = "1407",
  year          = "1993"
}

@article{MMV,
  author        = "McGough, Lauren and Mezei, M{\'a}rk and Verlinde, Herman",
  title         = "{Moving the CFT into the bulk with $T\bar T$}",
  eprint        = "1611.03470",
  archivePrefix = "arXiv",
  primaryClass  = "hep-th",
  journal       = "JHEP",
  volume        = "04",
  pages         = "010",
  year          = "2018"
}

@article{KLM,
  author        = "Kraus, Per and Liu, Junyu and Marolf, Donald",
  title         = "{Cutoff AdS$_3$ versus the $T\bar T$ deformation}",
  eprint        = "1801.02714",
  archivePrefix = "arXiv",
  primaryClass  = "hep-th",
  journal       = "JHEP",
  volume        = "07",
  pages         = "027",
  year          = "2018",
  doi           = "10.1007/JHEP07(2018)027"
}

@article{HKST,
  author        = "Hartman, Thomas and Kruthoff, Jorrit and Shaghoulian, Edgar and Tajdini, Amirhossein",
  title         = "{Holography at finite cutoff with a $T^2$ deformation}",
  eprint        = "1807.11401",
  archivePrefix = "arXiv",
  primaryClass  = "hep-th",
  journal       = "JHEP",
  volume        = "03",
  pages         = "004",
  year          = "2019"
}

@article{StanfordYang,
  author        = "Stanford, Douglas and Yang, Zhenbin",
  title         = "{Finite-cutoff JT gravity and self-avoiding loops}",
  eprint        = "2004.08005",
  archivePrefix = "arXiv",
  primaryClass  = "hep-th",
  year          = "2020"
}

@article{AnninosHarris,
  author        = "Anninos, Dionysios and Harris, Eleanor",
  title         = "{Three-dimensional de Sitter horizon thermodynamics}",
  eprint        = "2106.13832",
  archivePrefix = "arXiv",
  primaryClass  = "hep-th",
  journal       = "JHEP",
  volume        = "10",
  pages         = "091",
  year          = "2021"
}

@article{Coleman,
  author        = "Coleman, Evan and Mazenc, Edward A. and Shyam, Vasudev and Silverstein, Eva and Soni, Ronak M. and Torroba, Gonzalo and Yang, Sungyeon",
  title         = "{De Sitter microstates from $T\bar T+\Lambda_2$ and the Hawking-Page transition}",
  eprint        = "2110.14670",
  archivePrefix = "arXiv",
  primaryClass  = "hep-th",
  journal       = "JHEP",
  volume        = "07",
  pages         = "140",
  year          = "2022",
  doi           = "10.1007/JHEP07(2022)140"
}

@article{BanihashemiJacobson,
  author        = "Banihashemi, Batoul and Jacobson, Ted",
  title         = "{Thermodynamic ensembles with cosmological horizons}",
  eprint        = "2204.05324",
  archivePrefix = "arXiv",
  primaryClass  = "hep-th",
  journal       = "JHEP",
  volume        = "07",
  pages         = "042",
  year          = "2022"
}

@article{SVVV,
  author        = "Svesko, Andrew and Verheijden, Evita and Verlinde, Erik P. and Visser, Manus R.",
  title         = "{Quasi-local energy and microcanonical entropy in two-dimensional nearly de Sitter gravity}",
  eprint        = "2203.00700",
  archivePrefix = "arXiv",
  primaryClass  = "hep-th",
  journal       = "JHEP",
  volume        = "08",
  pages         = "075",
  year          = "2022"
}

@article{AndradeDirichlet,
  author        = "Andrade, Tom{\'a}s and Kelly, William R. and Marolf, Donald and Santos, Jorge E.",
  title         = "{On the stability of gravity with Dirichlet walls}",
  eprint        = "1504.07580",
  archivePrefix = "arXiv",
  primaryClass  = "gr-qc",
  journal       = "Class. Quant. Grav.",
  volume        = "32",
  pages         = "235006",
  year          = "2015"
}

@article{Anderson08,
  author        = "Anderson, Michael T.",
  title         = "{On boundary value problems for Einstein metrics}",
  eprint        = "math/0612647",
  archivePrefix = "arXiv",
  journal       = "Geom. Topol.",
  volume        = "12",
  pages         = "2009",
  year          = "2008"
}

@article{WittenBC,
  author        = "Witten, Edward",
  title         = "{A note on boundary conditions in Euclidean gravity}",
  eprint        = "1805.11559",
  archivePrefix = "arXiv",
  primaryClass  = "hep-th",
  journal       = "Rev. Math. Phys.",
  volume        = "33",
  pages         = "2140004",
  year          = "2021"
}

@article{AnAnderson21,
  author        = "An, Zhongshan and Anderson, Michael T.",
  title         = "{The initial boundary value problem and quasi-local Hamiltonians in general relativity}",
  eprint        = "2103.15673",
  archivePrefix = "arXiv",
  primaryClass  = "gr-qc",
  journal       = "Class. Quant. Grav.",
  volume        = "38",
  pages         = "154001",
  year          = "2021"
}

@article{AnAnderson25,
  author        = "An, Zhongshan and Anderson, Michael T.",
  title         = "{Well-posed geometric boundary data in general relativity, III: Conformal-mean curvature boundary data}",
  eprint        = "2503.12599",
  archivePrefix = "arXiv",
  primaryClass  = "math.AP",
  year          = "2025"
}

@article{AGMflat,
  author        = "Anninos, Dionysios and Galante, Dami{\'a}n A. and Maneerat, Chawakorn",
  title         = "{Gravitational observatories}",
  eprint        = "2310.08648",
  archivePrefix = "arXiv",
  primaryClass  = "hep-th",
  journal       = "JHEP",
  volume        = "12",
  pages         = "024",
  year          = "2023"
}

@article{AAGMAdS4,
  author        = "Anninos, Dionysios and Arias, Ra{\'u}l and Galante, Dami{\'a}n A. and Maneerat, Chawakorn",
  title         = "{Gravitational observatories in AdS$_4$}",
  eprint        = "2412.16305",
  archivePrefix = "arXiv",
  primaryClass  = "hep-th",
  journal       = "JHEP",
  volume        = "07",
  pages         = "234",
  year          = "2025",
  doi           = "10.1007/JHEP07(2025)234"
}

@article{LSW,
  author        = "Liu, Xiaoyi and Santos, Jorge E. and Wiseman, Toby",
  title         = "{New well-posed boundary conditions for semi-classical Euclidean gravity}",
  eprint        = "2402.04308",
  archivePrefix = "arXiv",
  primaryClass  = "hep-th",
  journal       = "JHEP",
  volume        = "06",
  pages         = "044",
  year          = "2024"
}

@article{BSSflat,
  author        = "Banihashemi, Batoul and Shaghoulian, Edgar and Shashi, Sanjit",
  title         = "{Flat space gravity at finite cutoff}",
  eprint        = "2409.07643",
  archivePrefix = "arXiv",
  primaryClass  = "hep-th",
  journal       = "Class. Quant. Grav.",
  volume        = "42",
  pages         = "035010",
  year          = "2025"
}

@article{BSSthermal,
  author        = "Banihashemi, Batoul and Shaghoulian, Edgar and Shashi, Sanjit",
  title         = "{Thermal effective actions from conformal boundary conditions in gravity}",
  eprint        = "2503.17471",
  archivePrefix = "arXiv",
  primaryClass  = "hep-th",
  journal       = "Class. Quant. Grav.",
  volume        = "42",
  pages         = "155004",
  year          = "2025",
  doi           = "10.1088/1361-6382/adee72"
}

@article{Caceres:2026yvs,
    author = "C{\'a}ceres, Elena and Krishna, Hare and Balaji, Harita Palani and Patil, Vaishnavi",
    title = "{Holographic entanglement entropy with conformal boundary conditions}",
    eprint = "2608.06277",
    archivePrefix = "arXiv",
    primaryClass = "hep-th",
    reportNumber = "UT-WI-21-2026",
    month = "8",
    year = "2026"
}

@article{LiuReallSantosWiseman25,
    author        = "Liu, Xiaoyi and Reall, Harvey S. and Santos, Jorge E. and Wiseman, Toby",
    title         = "{Ill-posedness of the Cauchy problem for linearized gravity in a cavity with conformal boundary conditions}",
    eprint        = "2505.20410",
    archivePrefix = "arXiv",
    primaryClass  = "gr-qc",
    doi           = "10.1088/1361-6382/ae1b60",
    journal       = "Class. Quant. Grav.",
    volume        = "42",
    number        = "23",
    pages         = "235003",
    year          = "2025"
}

@article{Ghosh:2019rcj,
    author = "Ghosh, Animik and Maxfield, Henry and Turiaci, Gustavo J.",
    title = "{A universal Schwarzian sector in two-dimensional conformal field theories}",
    eprint = "1912.07654",
    archivePrefix = "arXiv",
    primaryClass = "hep-th",
    doi = "10.1007/JHEP05(2020)104",
    journal = "JHEP",
    volume = "05",
    pages = "104",
    year = "2020"
}

@article{ChoiMezeiSarosi,
  author        = {Choi, Changha and Mezei, M{\'a}rk and S{\'a}rosi, G{\'a}bor},
  title         = {Exact four point function for large $q$ {SYK} from {Regge} theory},
  journal       = {JHEP},
  volume        = {05},
  pages         = {166},
  year          = {2021},
  doi           = {10.1007/JHEP05(2021)166},
  eprint        = {1912.00004},
  archivePrefix = {arXiv},
  primaryClass  = {hep-th}
}

@article{BTZ,
  author        = "Ba{\~n}ados, M{\'a}ximo and Teitelboim, Claudio and Zanelli, Jorge",
  title         = "{The black hole in three-dimensional space-time}",
  eprint        = "hep-th/9204099",
  archivePrefix = "arXiv",
  journal       = "Phys. Rev. Lett.",
  volume        = "69",
  pages         = "1849",
  year          = "1992",
  doi           = "10.1103/PhysRevLett.69.1849"
}
